\documentclass[prc,twocolumn,nofootinbib,tightenlines,superscriptaddress,showpacs,preprintnumbers]{revtex4}

\usepackage{amsmath,amssymb,amsfonts,graphicx,subfigure,wrapfig,slashed}

\newcommand{\be}{\begin{equation}}
\newcommand{\ee}{\end{equation}}

\def\hs{\hspace}
\def\no{\nonumber}

\begin{document}

\title{Analytic approach to nuclear rotational states: The role of spin \\ 
 - A minimal model - }

\author{W. Bentz}
\email[Corresponding author:~]{bentz@keyaki.cc.u-tokai.ac.jp}
\affiliation{Department of Physics, School of Science, Tokai University,
             4-1-1 Kitakaname, Hiratsuka-shi, Kanagawa 259-1292, Japan}

\author{A. Arima}
\affiliation{Musashi University,
1-26-1 Toyotama-kami, Nerima-ku, Tokyo 176-8534, Japan}


\author{A. Richter}
\affiliation{Institut f\"ur Kernphysik, Technische Universit\"at Darmstadt,
Schlossgartenstrasse 9, D-64289 Darmstadt, Germany}

\author{J. Wambach}
\affiliation{Institut f\"ur Kernphysik, Technische Universit\"at Darmstadt,
Schlossgartenstrasse 9, D-64289 Darmstadt, Germany}

\begin{abstract}
\begin{description}
\item[Background:] 
The scissors mode is a rotational mode of isovector character in deformed nuclei.
Together with the isoscalar rotation, it is the prominent collective
mode at low excitation energies.  
\item[Purpose:] We use a simple field theory model to investigate the role 
of the nucleon spin for the magnetic sum rules associated with the low-lying scissors 
mode. Special emphasis is put on the coupling of the spin 
part of the M1 operator to the scissors mode.
\item[Methods:] We apply the mean field approximation and random phase approximation (RPA) 
to a model Hamiltonian based on a simple quadrupole-quadrupole interaction. The effects of
the spin-orbit interaction are included in the mean field Hamiltonian.
Ward-Takahashi relations are used to derive the M1 sum rules.
\item[Results:] 
The presence of the spin-orbit interaction leads to  
interference terms in the inverse energy weighted sum rule. It is shown that
the low-lying scissors mode, which is generated by the isovector combination
of proton and neutron total angular momenta, gives the main contribution to
the spectral sum for this case.  
\item[Conclusions:] 
The basic concept of the scissors mode as an isovector
vibrational rotation remains valid in the presence of the nucleon spin.
The inverse energy weighted sum rule, however, receives non-trivial modifications 
because of interference terms.  
\end{description}
\end{abstract}

\pacs{21.10.Ky,21.10.Re,21.60.Ev}

\maketitle

\section{Introduction}
\setcounter{equation}{0}

The scissors mode, which is a rotational mode of isovector character in
deformed nuclei\cite{SC,NCIM}, is usually visualized by an out of phase orbital motion
of protons against neutrons. This intuitive picture lead to the prediction of the 
scissors mode both in terms of
nucleonic\cite{TWOROT} and bosonic\cite{FI} degrees of freedom, and has formed the 
basis of many investigations \cite{SCREV,IU1,ZA}. In some respect, the scissors mode
can be thought of as an extension of the familiar isoscalar collective rotation\cite{BM} 
to the isovector rotational motion. 
Indeed, in a recent work \cite{BAERW} we have used a simple field theory model 
for the collective orbital motion to 
show how an effective Hamiltonian of the 2-rotor type emerges 
naturally from the cooperation of the isoscalar (Goldstone) and
isovector (scissors) degrees of freedom.  
 
In usual nuclear Hamiltonians, the orbital angular momentum alone is 
not conserved. A familiar example is the spin-orbit potential, which explicitly
breaks the separate spin and orbital symmetries on the level of the mean field
Hamiltonian. The explicit breaking of the spin symmetry on the mean field level 
has important consequences for the properties of nuclear rotational states, 
and we mention two of them here in more detail: First, if we use the mechanism 
of spontaneous symmetry breaking to generate a deformed mean field,  
it is the {\em total} angular momentum symmetry which will be restored
by the collective rotation of the whole system. That is, the isoscalar Goldstone
modes are generated essentially by the total angular momentum 
${\bold J}={\bold J}_p + {\bold J}_n$, and their 
degrees of freedom will lead to an effective rotational Hamiltonian of the schematic 
form ${\bold J}^2/(2 I^J)$, where $I^J=I^J_p + I^J_n$ is the total moment of inertia
\footnote{For a discussion on the connection between Goldstone modes and the emergence
of rotational bands in finite systems, see e.g., Ref.\cite{UTFU}. Recently, more general
and extensive 
discussions have been presented in the framework of effecive field theories \cite{PW}.}. 
In analogy to the situation found for the pure orbital
case\cite{BAERW,SRM}, we can then anticipate the existence of a mode at low excitation
energy, which is generated by the isovector combination of ${\bold J}_p$ and ${\bold J}_n$, 
and which
cooperates with the Goldstone mode so as to give a total effective 
rotational Hamiltonian of the schematic 2-rotor
form ${\bold J}_p^2/(2 I_p^J) + {\bold J}_n^2/(2 I_n^J)$. 
The structure of the Goldstone and scissors modes can therefore be expected to 
come out similar 
to the case of the pure orbital motion, but with the orbital angular momenta
replaced by the total angular momenta.  

Second, the explicit breaking of the spin symmetry on the mean field level
leads to new types of correlation functions, because the spin operators
(${\bold S}_p$ and ${\bold S}_n$) can generate particle-hole states of finite excitation energy.
These new types of non-interacting correlators, which are visualized by the bubble graphs
shown by Fig. 2 of the next section with one of the external operators 
being a spin operator, will obviously modify the sum rules, because the latter ones
can be represented in terms of correlation functions\cite{LS,PRC57}. In particular,
we will see that the presence of these correlators implies that the 
spin part of the M1 operator can couple to the low energy scissors
modes. This observation is of particular interest, because it is known that there 
exists a discrepancy between the theoretical and experimental values of the
inverse energy weighted M1 sum rule, for which the low energy scissors mode is expected
to give the dominant contribution\cite{PRC71,BAERW}. 
 
The purpose of the present paper is to consider a ``minimal model'' to
get analytic insight into the role of the spin for the low energy 
rotational states in deformed nuclei. By ``minimal'' we mean that the spin dependence
of the nuclear Hamiltonian is taken into account only via the one-body 
spin-orbit potential, and not via the two-body interactions. 
Besides simplicity, the reason to restrict ourselves to this ``minimal model''
is to keep the discussions distinct from the collective spin modes
at higher energies, which are known to be strongly modified by interactions of the spin-spin
type\cite{ZMS}.
In order to concentrate on the important role of the spin-orbit potential, we also
will not include the effects of pairing correlations in this paper.
The role of spin dependent interactions and pairing correlations for the M1 sum
rules will be discussed in a separate work\cite{NEW}.   
 
 

In Sect. II we will briefly explain the model and the approximations (mean field
approximation and random phase approximation (RPA)). 
In Sect. III we will explain the constraints which follow from angular momentum
conservation, and in Sect. IV we will discuss the M1 sum rules, with a special
emphasis on the role of the spin in the inverse energy weighted (IEW) sum rule. 
Sect. V gives a summary of our results. More formal discussions and derivations
are presented in four Appendices.

\section{Model and approximation schemes}
\setcounter{equation}{0}

Our model is based on the simple quadrupole-quadrupole ($QQ$) interaction, which we
used in our earlier work\cite{BAERW} to describe the pure orbital collective motion. 
Because the mean field approximation and RPA for this model 
are discussed in detail in that paper, we only
summarize the main points here. 

The one-body part of the Hamiltonian contains a spherical mean field $U_0(x)$, 
which we will not need to specify, and a spin-orbit interaction of the simplest
possible form from (see Eq.(\ref{h0}) below). 
The two-body interaction, which we need to describe the collective rotational states, 
is of the form 
\begin{align}
V_{QQ} = \sum_{\tau \rho} \frac{\chi_{\tau \rho}}{2} 
Q^{\dagger}_{\tau} \cdot Q_{\rho} \,, 
\label{vqq}
\end{align}
where $\tau$ denotes protons ($p$) and neutrons ($n$), and for the coupling constants we
assume $\chi_{pp}=\chi_{nn}$ and $\chi_{pn}=\chi_{np}$. The quadrupole
operator is defined by
\begin{align}
Q_{\tau}^{K} = \int {\rm d}^3 x  \, \psi^{\dagger}_{\tau}(x) Q^{K}(x) \psi_{\tau}(x) 
\,,   \label{q}
\end{align}
where $Q^{K}(x) = r^2 Y_{2 K}(\hat{x})$. Because the spin-orbit potential breaks the
spin symmetry explicitly from the outset, the Hamiltonian is only invariant
under rotations generated by the total angular momentum ${\bold J} = {\bold L}
+ {\bold S}$.   

The assumption that the $K=0$ component of the quadrupole
operator has a finite ground-state expectation value
\begin{align}
Q_{\tau}^0 = \langle Q_{\tau}^0 \rangle + : Q_{\tau}^0: \, \, 
\label{mfa}
\end{align} 
leads to spontaneous breaking of the rotational symmetry down to rotations generated
by $J^z=J^0$.    
The first term in (\ref{mfa}) expresses a deformed mean field for protons and neutrons, and
the dots in the second term denote normal ordering. By requiring self consistency for
the strength of the deformed mean fields, one finally arrives at the
following form of the Hamiltonian \cite{BAERW}: 
\begin{align}
H = H_{0p} + H_{0n} + C   
+ \sum_{\tau \rho} \frac{\chi_{\tau \rho}}{2} \left(:Q^{\dagger}_{\tau}:\right) 
\cdot \left(:Q_{\rho}:\right)\,,   \label{hfin}
\end{align}
where the one-body parts are given by
\begin{multline}
H_{0\tau}  
=  \int {\rm d}^3 x  \\
\times 
\psi_{\tau}^{\dagger}(x) \left( - \frac{\Delta}{2M} +
U_0(x) - \gamma_{\tau} {\bold L} \cdot {\bold S} - \beta_{\tau} Q^0(x) \right) \psi_{\tau}(x)
\label{h0}
\end{multline}
and $C$ is a constant given by 
\begin{align}
C &= - \frac{\chi_{pp}}{2} \left( \langle Q_{p}^0 \rangle^2 + 
\langle Q_{n}^0 \rangle^2 \right) - \chi_{pn} \langle Q_{p}^0 \rangle \langle Q_{n}^0 \rangle.
\label{c}
\end{align}
The parameters $\gamma_{\tau}$ in (\ref{h0}) represent the strength of the spin-orbit
potential. The deformation parameters $\beta_{\tau}$ obey the self consistency relations
\begin{align}
\beta_{\tau} &= \varepsilon_{\tau} -\sum_{\lambda} \chi_{\tau \lambda}  \langle Q_{\lambda}^0 \rangle
\,\,,
\label{gap} 
\end{align}
where the parameters $\varepsilon_{\tau}$, which describe the explicit breaking of the
rotational symmetry via a term $-\sum_{\tau} \varepsilon_{\tau}Q_{\tau}^0$ in the Hamiltonian, 
are introduced only for technical reasons and will be set to
zero in all final results.  
 
The excitation energies and wave functions of the collective states of the system
are determined by the poles and the residues of particle-hole T-matrix, which is a solution
of the Bethe-Salpeter (BS) equation, or equivalently the RPA equation. Because we are
assuming axial symmetry, we can characterize the states by the quantum number
$K$, which is the projection of the total angular momentum on the symmetry ($z$) 
axis. Since we are interested in the isoscalar Goldstone and the isovector scissors
rotational states, we consider the BS equation in the $K=1$ channel, which is
graphically represented by Fig. 1. 
\begin{figure}[tbp]
\begin{center}
\includegraphics[width=\columnwidth]{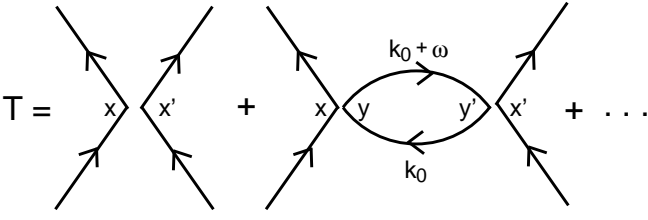}
\caption{Graphical representation of the particle-hole T-matrix in the ladder
approximation. For the explicit form of the equation, see Eq.(10) of Ref.\cite{BAERW}.
The total energy in the particle-hole channel ($\omega$) and the positions
$\vec{x}$, $\vec{x'}$ are fixed, and an integral is taken over the other
variables.}    
\end{center}
\end{figure}
Its solution is of the following form:
\begin{align}
T_{\tau \rho}  \left({x}', {x}; \omega\right)
\equiv - i Q^1_{\tau}({x}') \, t_{\tau \rho}(\omega) \, Q^{1\dagger}_{\rho}({x}) \,,
\label{t}
\end{align}
where the reduced T-matrix satisfies the equation 
\begin{align}
t(\omega) &= \chi - \chi \, \pi^{QQ}(\omega) \, t(\omega) \no  \\
\Rightarrow 
t(\omega) &= \frac{1}{1 + \chi \pi^{QQ}(\omega)} \, \chi
= \chi \, \frac{1}{1 + \pi^{QQ}(\omega) \chi}.   \label{t1} 
\end{align}
Here the quantities $t_{\tau \rho}$, the coupling constants $\chi_{\tau \rho}$ and the
non-interacting polarizations (bubble graphs) with two external quadrupole
operators $\pi_{\tau \rho}^{QQ} = \delta_{\tau \rho}
\pi_{\tau}^{QQ}$ are considered as $2 \times 2$ matrices in charge space.
Having in mind later applications, we define the proton and neutron bubble 
graphs for arbitrary external
tensor operators ($K=1$ components $W'^1$ and $W^1$) as follows
(see Fig.2):
\begin{figure}[tbp]
\begin{center}
\includegraphics[width=\columnwidth]{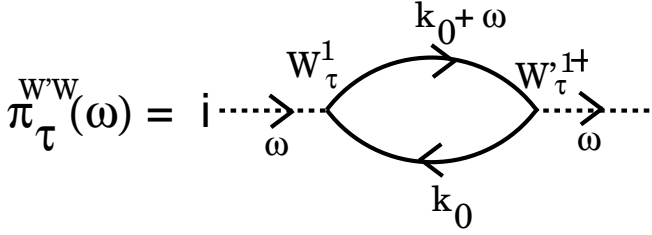}
\caption{Graphical representation of the bubble graph Eq.(\ref{pi}).}
\end{center}
\end{figure}
\begin{align}
\pi^{W'W}_{\tau}(\omega) 
&= i \int \frac{{\rm d}k_0}{2\pi} \int {\rm d}^3 x \int {\rm d}^3 x' \no \\
&\times
\left[ W'^{1\dagger}_{\tau}({x}') S_{\tau}\left({x}', {x}; k_0 + \omega \right)
W_{\tau}^1({x}) S_{\tau}\left({x}, {x}'; k_0 \right) \right]
\label{piprop} \\
&= - 2 \sum_{(\alpha i) \in \tau}
\langle \alpha | W'^1 | i \rangle^*  \langle \alpha | W^1 | i \rangle 
\frac{\Omega}{\omega^2 - \omega_{\alpha i}^2 + i \delta} \,.
\label{pi}
\end{align}
Here 
$S(x', x; k_0)$ denotes the Feynman propagator in the mixed representation,
$\omega_{\alpha i} = \epsilon_{\alpha} - \epsilon_i$ are the non-interacting 
particle-hole energies for protons (case $(\alpha i) \in p$) or neutrons
(case $(\alpha i) \in n$), and the symbol $\Omega$ means   
$\Omega=\omega_{\alpha i}$ if the operators $W$ and $W'$ have the same time reversal ($T$)
symmetry (case $t_W=t_{W'}$), and $\Omega=\omega$ if they have the opposite
$T$-symmetry ($t_W = - t_{W'}$)
\footnote{In the phase convention used in this paper, the $T$-symmetry of an operator 
${\cal O}$ is defined as in 
Eq.(A.22) of \cite{RO} by  
${\displaystyle \langle \overline{i} | {\cal O} | \overline{\alpha} \rangle 
= t_{{\cal O}} \langle \alpha | {\cal O} | i \rangle}$, where the time-reversed
single particle states $\left(\overline{\alpha} \overline{i}\right)$ have the
opposite values of $j_z=\ell_z+s_z$ but the same energies as $\left(\alpha i \right)$
for the case of axial symmetry. The sign is $t_{\cal O} = +1$ for a $T$-even operator 
(like the quadrupole operator), 
and $t_{\cal O} = -1$ for a $T$-odd operator (like the angular momentum operators).}.

From (\ref{t1}), the poles of the T-matrix ($\omega^2 \equiv \omega_n^2$) are 
determined by the equation
\begin{align}
&{\rm Det} \left(1 + \chi \pi^{QQ}(\omega)\right) =0  \,,
\label{det}
\end{align}
and the reduced vertex functions for the collective states ($N_{\tau}(\omega_n$))
are defined by the residues of the reduced T-matrix:
\begin{align}
t_{\tau \rho}(\omega) &\stackrel{\omega^2 \rightarrow \omega_n^2}{\longrightarrow}
\frac{N_{\tau}(\omega_n) N_{\rho}(\omega_n)}{\omega^2 - \omega_n^2 + i \delta}. 
\label{pole}
\end{align}
From this definition and Eq.(\ref{t1}) one obtains the homogeneous BS equation and the
normalization condition for the reduced vertex functions:
\begin{align}
N_{\tau}(\omega_n) + \sum_{\lambda} \chi_{\tau \lambda} \pi^{QQ}_{\lambda}(\omega_n) 
N_{\lambda}(\omega_n) 
&=0  \label{hombs}   \\
N_p^2(\omega_n) \, \pi_p^{QQ}{}'(\omega_n)   \, 
+ N_n^2(\omega_n) \,  \pi_n^{QQ}{}'(\omega_n) &= 1 \,.    
\label{norm}
\end{align}
The prime in (\ref{norm}) indicates differentiation w.r.t. $\omega^2$, i.e.,
%
\begin{align}
\pi^{QQ}_{\tau}{}'(\omega) &\equiv \frac{{\rm d}\pi_{\tau}^{QQ}}{{\rm d}\omega^2}
\no \\
&= 2 \sum_{(\alpha i) \in \tau}
|\langle \alpha | Q_{\tau}^1 | i \rangle |^2  
\frac{\omega_{\alpha i}}{\left(\omega^2 - \omega_{\alpha i}^2 + i \delta \right)^2} \,.
\label{pip}
\end{align}
The full vertex functions $\Gamma_{\tau}^n(x)$ for the $K=1$ collective states
with excitation energy $\omega_n$ are defined similarly to (\ref{pole}) by
the residues of the full T-matrix 
(\ref{t}), and are given in the present model by
\begin{align}
\Gamma_{\tau}^n(x) = Q^1_{\tau}(x) \, N_{\tau}(\omega_n) \,. \label{fullv} 
\end{align}
The relation of this BS formalism to the usual RPA
is explained in Appendix A, where it is shown that (\ref{norm}) is
equivalent to the familiar normalization condition for the forward and backward
RPA amplitudes.

\section{Constraints from rotational symmetry}
\setcounter{equation}{0}

By definition, a tensor operator $W^q$ of rank $k$ with spherical components
$q=-k, \dots, k$ satisfies the following commutations relations with the total angular
momentum operators\cite{MES} 
\footnote{In this paper, we use the definitions $V^{\pm 1} = \mp (V^x \pm i V^y)/\sqrt{2}$,
$\,V^0=V^z$ for all vector operators $V$, including
the case of the angular momentum operators ($V=L,\, S,\, J$). Therefore the sign of the
$q=1$ component of the angular momentum operators is opposite to 
Ref.\cite{MES} or \cite{BAERW}.} 
$\, {\displaystyle J^{\pm 1} = \mp \tfrac{1}{\sqrt{2}}\left(J^x \pm i J^y
\right)}$, ${\displaystyle J^0 = J^z}$\,:
\begin{align}
\left[ J^{\pm 1}, W^q \right] &= \mp \frac{1}{\sqrt{2}}
\sqrt{k(k+1) - q (q \pm 1)} \, W^{q \pm 1}    \label{comm1}
\\\
\left[ J^0 , W^q \right] &= q \, W^q \label{comm2}  \,.
\end{align}
The commutator of $H_{0\tau}$ (Eq.(\ref{h0})) with
total angular momentum operators then becomes
\begin{align}
\left[ H_{0 \tau}, J_{\tau}^{\pm 1} \right] = \mp \sqrt{3} \beta_{\tau}
Q_{\tau}^{\pm 1}\,, \,\,\,\,\,\,\left[ H_{0 \tau}, J_{\tau}^{0} \right] = 0 \,,
\label{comm0}
\end{align}
and if we consider matrix elements of these identities between non-interacting 
particle-hole states, we obtain the useful relations
\begin{align}
\langle \alpha | J_{\tau}^{\pm 1} | i \rangle
= \mp \frac{\sqrt{3} \beta_{\tau}}{\omega_{\alpha i}}
\langle \alpha | Q_{\tau}^{\pm 1} | i \rangle \,, \,\,\,\,\,
\langle \alpha | J_{\tau}^{0} | i \rangle = 0\,.
\label{rel}
\end{align}
Here $\left(\alpha i\right)\in \tau$, and the above relations hold for non-degenerate
particle-hole states ($\omega_{\alpha i} \neq 0$). 
Let us note here two applications of the relations (\ref{rel}):
First, consider the bubble graph (\ref{pi}) for the case where $W'=Q$ is the quadrupole operator 
which is $T$-even, and another operator which is $T$-odd, for example one of the
angular momentum operators $R \equiv L, \, S$ or $J$. Eq. (\ref{rel}) for $K=1$ then leads to the
relation
\begin{align}
\pi_{\tau}^{Q R}(\omega) = - \frac{\omega}{\sqrt{3} \beta_{\tau}} \, 
\pi_{\tau}^{J R}(\omega)\,.  \label{rel1}
\end{align}
Second, the derivative of the $QQ$ bubble graph (\ref{pip}) at $\omega=0$ can be expressed in 
terms of the $JJ$ bubble graph as follows:
\begin{align}  
\pi^{QQ}_{\tau}{}'(0) 
= \frac{2}{3 \beta_{\tau}^2} 
\sum_{(\alpha i) \in \tau} \frac{
|\langle \alpha | J^1_{\tau} | i \rangle|^2} {\omega_{\alpha i}}
= \frac{1}{3 \beta_{\tau}^2} \, \pi^{JJ}_{\tau}(0) \,. 
\label{pip1}
\end{align}
The quantities $\pi_{\tau}^{JJ}(0)$ are actually the moments of inertia of protons or neutrons, 
according to the familiar Inglis formula\cite{ING}, expressed in terms of the total angular 
momentum operators:
\begin{align}
I^J_{\tau} \equiv \pi_{\tau}^{JJ}(0) &= 2 \sum_{(\alpha i) \in \tau} \frac{
|\langle \alpha | J^1_{\tau} | i \rangle|^2} {\omega_{\alpha i}} 
\label{ingpn} \\
I^J &\equiv I^J_p + I^J_n \,.  \label{ingtot}   
\end{align}
Here $I^J$ is the total moment of inertia. 

Next let us discuss the Ward-Takahashi identities\cite{YT} for the full $K=1$ 
correlation functions, which 
follow from angular momentum conservation. 
We consider the time derivative of the 2-point function 
${\displaystyle \langle 0| T \left( W_{\tau}^{1 \dagger}(t') \, J^1(t) \right) |0 \rangle}$
with one arbitrary tensor operator $W^1_{\tau}(t')$ of rank $k$ and the total 
angular momentum operator
$J^{1}(t) = J_p^{1}(t) + J_n^1(t)$. Using the Heisenberg equation of motion
\begin{align}
\frac{\partial J^{1}}{\partial t} = i \left[ H, J^{1} \right]
= - i \sqrt{3} \sum_{\tau} \varepsilon_{\tau} Q_{\tau}^{1},
\label{heis}
\end{align}
and the equal time commutator $\left[ J^1(t), W^{1\dagger}_{\tau}(t) \right]$
from (\ref{comm1}), and performing a Fourier transformation $(t' - t) \rightarrow \omega$, 
we obtain the Ward-Takahashi identity
\footnote{We use 
the symbol $\Pi^{W'W}_{\lambda \tau}$ for the exact correlators and the correlators
in the chain approximation (RPA), and $\pi^{W'W}_{\lambda \tau} = 
\delta_{\lambda \tau} \pi^{W'W}_{\tau}$ for the non-interacting ones. Although not
indicated explicitly, all correlators refer to the case $K=1$. To get the last term in 
Eq.(\ref{wt1}) we assumed the relation $\left(W^K\right)^{\dagger} = (-1)^K W^{-K}$,
which is satisfied for all operators of the main text. (Only for the operator $V$ defined
by (\ref{v}) the sign is opposite.)}  
\begin{align}
\sum_{\lambda} \, \omega \, \Pi_{\tau \lambda}^{WJ}(\omega) = - \sqrt{3}\, 
\sum_{\lambda} \Pi_{\tau \lambda}^{WQ} (\omega) \varepsilon_{\lambda} 
+ \sqrt{\frac{k(k+1)}{2}} \, \langle W^0_{\tau}  \rangle. 
\label{wt1}
\end{align}
Here we defined the {\em exact} 2-point functions
for arbitrary tensor operators $W'$ and $W$ by 
\begin{align}
\langle 0 |
T \left( W'^{1 \dagger}_{\tau}(t') \, W^1_{\lambda}(t) \right) |0 \rangle 
&\equiv -i \, \Pi_{\tau \lambda}^{W'W}(t'-t)   \no \\
&\hs{-7mm}
= -i \int {\rm d}\omega \, e^{- i \omega (t'-t)} \Pi_{\tau \lambda}^{W'W}(\omega).
\label{corr} 
\end{align}
To visualize these correlators in our chain approximation (RPA), 
we represent them graphically in Fig.3, which translates into the following expressions 
in terms of the reduced particle-hole t-matrix of
Eq.(\ref{t1}):
\begin{align}
\Pi_{\tau \lambda}^{W' W}(\omega) &= \delta_{\lambda \tau} 
\pi^{W'W}_{\tau}(\omega) - \sum_{\rho} \pi^{W'Q}_{\tau}(\omega) \chi_{\tau \rho}
\Pi^{QW}_{\rho \lambda}(\omega)  
\label{rpag1} \\
&= \delta_{\lambda \tau} 
\pi^{W'W}_{\tau}(\omega) - \pi^{W'Q}_{\tau}(\omega) t_{\tau \lambda}(\omega) 
\pi^{QW}_{\lambda}(\omega) \,.  
\label{rpag2}
\end{align} 
\begin{figure*}[tbp]
\begin{center}
\includegraphics[width=2\columnwidth]{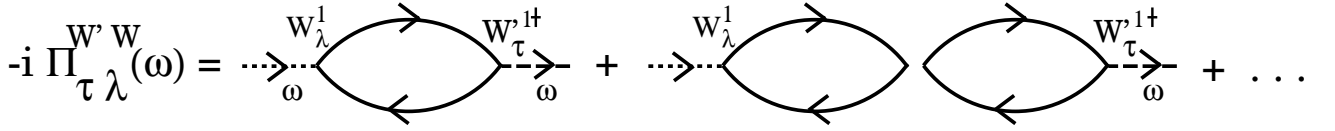}
\caption{Graphical representation of the 2-point function $\Pi^{W' W}_{\tau \lambda}$, 
Eq.(\ref{rpag1}).}
\end{center}
\end{figure*}  
For later reference, we also note that the spectral representation of the exact correlators 
is given in analogy to Eq.(\ref{pi}) for the non-interacting ones by
\begin{align}
&\Pi_{\tau \lambda}^{W'W}(\omega) \no \\
%
&= -2 \sum_n \langle n| W'^{1}_{\tau} | 0 \rangle ^* \langle n| W_{\lambda}^1 
|0 \rangle
\frac{\Omega}{\omega^2 - \Omega_n^2 + i \delta}\,, \label{spec1}
\end{align}
where $\Omega_n$ are the exact excitation energies of the $K=1$ eigenstates
$|n \rangle$ of the Hamiltonian $H$, and
$\Omega$ is defined as in Eq.(\ref{pi}) by $\Omega = \Omega_n$ (if $t_{W'}=t_W$)
and $\Omega = \omega$ (if $t_{W'} = - t_W$).

Returning to the Ward-Takahashi relation (\ref{wt1}), there are two important limits, 
namely the ``Goldstone limit'' ($\omega \rightarrow 0$ for finite $\varepsilon_{\lambda}$), 
and the ``exact symmetry limit'' ($\varepsilon_{\lambda} \rightarrow 0$ for finite $\omega$).
In the Goldstone limit we have
\begin{align}
\sum_{\lambda} \Pi_{\tau \lambda}^{WQ} (0) \, \varepsilon_{\lambda} 
= \sqrt{\frac{k(k+1)}{6}} \langle W^0_{\tau} \rangle\,, \label{w1}
\end{align} 
while in the exact symmetry limit we have
\begin{align}
\omega \, \sum_{\lambda} \Pi_{\tau \lambda}^{WJ} (\omega) \, 
= \sqrt{\frac{k(k+1)}{2}} \langle W^0_{\tau} \rangle\,. \label{w2}
\end{align} 
Each of these limits has important implications. Consider first the Goldstone limit (\ref{w1})
for the case where $W$ is $T$-odd. We know from the spectral representation Eq.(\ref{spec1}) 
that the l.h.s. of (\ref{w1}) then vanishes, which implies that the ground state 
expectation value of the $K=0$ component of
any $T$-odd tensor operator is zero: $\langle W_{\tau}^0 \rangle = 0$ if $t_{W}=-1$. 
Using this result in the exact symmetry limit (\ref{w2}), we obtain the constraint
\begin{align}
\sum_{\lambda} \Pi_{\tau \lambda}^{W J}(\omega) = 0 \,\,\,\,\,\,\,\,\,\,\,\,
(\varepsilon_{\lambda}=0 \,,\,\,t_{W}=-1\,,\,\, k>0) \,.
\label{ang}
\end{align}
Later, in connection with the M1 sum rules, we will verify Eq.(\ref{ang}) explicitly for 
the case where $W$ is an angular momentum operator and $\omega=0$. 

The relation (\ref{ang}) holds if the tensor operator $W$ is $T$-odd. If $W$ is 
$T$-even and its $K=0$ component has a 
non-vanishing ground state expectation value, the rotational symmetry is spontaneously
broken, and $\sum_{\lambda} \Pi^{WJ}_{\tau \lambda}(\omega)$ is nonzero. The spectral
representation (\ref{spec1}) shows that only the Goldstone mode ($\Omega_0=0$) contributes
here, because the total angular momentum ${\bold J} = \sum_{\lambda} {\bold J}_{\lambda}$ 
cannot give rise to 
finite energy excitations because of $\left[H, {\bold J} \right]=0$.  
Then it follows from the spectral representation (\ref{spec1}) that for this case   
$\sum_{\lambda} \Pi^{WJ}_{\tau \lambda}(\omega) = \frac{K}{\omega}$, where $K$ is a constant,
in agreement with (\ref{w2}).

The Goldstone limit (\ref{w1}) for the case $W=Q$ can be used to confirm the existence of 
2 independent Goldstone modes, corresponding to the 2 broken generators $J^x$ and $J^y$, and to
determine the vertex functions of those modes. The arguments are very
similar to those given in Ref.\cite{BAERW}, and we briefly summarize the main points
here. First, to show the existence of the Goldstone modes, we note that the Dyson
equation (\ref{rpag1}) for the case $W'=W=Q$ takes the matrix form
${\displaystyle \left(1+\pi^{QQ}(\omega)\chi \right) \Pi^{QQ}(\omega) = \pi^{QQ}(\omega)}$.
Multiplying then (\ref{w1}) for $W=Q$ from left by 
$\left(1 + \pi^{QQ}(0) \chi \right)$ we obtain
\begin{align}
\pi^{QQ}(0) \left(\beta + \chi \langle Q^0 \rangle \right) = \left(1 + \pi^{QQ}(0) \chi
\right) \langle Q^0 \rangle ,  \nonumber
\end{align}
where we used the self consistency relation (\ref{gap}) to eliminate 
$\varepsilon$. 
(In this notation, $\beta$ and $\langle Q^0 \rangle$ are considered as vectors in charge space.)
We then obtain the identity
\begin{align}
\pi^{QQ}_{\tau}(0) \beta_{\tau} = \langle Q^0_{\tau} \rangle \,,  
\label{let}
\end{align}
by which the self consistency relation (\ref{gap}) can be rewritten as
\begin{align}
\beta_{\tau} = \varepsilon_{\tau} - \sum_{\lambda} \left(\chi_{\tau \lambda} 
\pi^{QQ}_{\lambda}(0)\right) \beta_{\lambda} .  \label{gap1}
\end{align}
In the limit of exact rotational symmetry ($\varepsilon_{\tau}=0$), 
this equation leads to the condition 
${\displaystyle {\rm Det} \left(1 + \chi \pi^{QQ}(0)\right) = 0}$ for a 
nontrivial solution. Comparing this with the
pole equation (\ref{det}), we see that in the limit of exact rotational symmetry 
the self consistency relation guarantees the existence of a Goldstone
pole ($\omega_0=0$) in the $K=1$ channel. A similar argument holds for the case $K=-1$,
which completes the proof for the existence of 2 Goldstone modes. The reduced
T-matrix (\ref{pole}) therefore behaves in the exact symmetry limit
($\varepsilon_{\lambda}=0$) as follows:
\begin{align}
t_{\tau \rho}(\omega) = \frac{N_{\tau}(0) N_{\rho}(0)}{\omega^2 + i \delta} +
\left({\rm terms} \,\,{\rm regular} \,\,{\rm for}\,\,\omega\rightarrow 0 \right) \,.
\label{sing}
\end{align}
Second, to determine the vertex functions of the Goldstone modes, we note that
a comparison of the homogeneous BS equation (\ref{hombs}) with (\ref{gap1}) 
in the limit of exact symmetry 
($\varepsilon_{\lambda}=0$) allows to specify the 
charge dependence of the Goldstone vertex functions as
\begin{align}
N_{\tau}(0) = N \, \beta_{\tau} \,,  \label{n0}
\end{align} 
where the charge independent constant $N$ is determined from the normalization 
condition (\ref{norm})
as 
\begin{align}
N = \left(\sum_{\tau} \beta_{\tau}^2 \pi^{QQ}_{\tau}{}'(0) \right)^{- \frac{1}{2}}
= \sqrt{\frac{3}{I^J}}\,,  \label{nn}
\end{align}
where we used the relations (\ref{pip1}) - (\ref{ingtot}). 
The form of the reduced and full $K=1$ 
Goldstone vertex functions (see Eq.(\ref{fullv})) can then be summarized as
\begin{align}
N_{\tau}(0) &= \sqrt{\frac{3}{I^J}} \,\beta_{\tau}  \label{rgv} \\
\Gamma_{\tau}^{n=0}({x}) &= \sqrt{\frac{3}{I^J}} \, Q^1_{\tau}({x})\, \beta_{\tau} .
\label{fgv}
\end{align}
The vertex function for the $K=-1$ Goldstone mode is obtained from (\ref{fgv}) by
replacing $Q^1 \rightarrow Q^{-1}$.  

Next we turn to the exact symmetry limit (\ref{w2}) for
the case $W=Q$. Since for the case of spontaneously broken symmetry the r.h.s. of 
this relation is a non-zero constant, we obtain in the limit $\omega \rightarrow \omega_n$,
where $\omega_n \neq 0$ is one of the nonzero solutions of the eigenvalue equation,
\begin{align}
\lim_{\omega^2 \rightarrow \omega_n^2} \, 
\left(\omega^2 - \omega_n^2 \right) \sum_{\lambda} \, 
\Pi_{\tau \lambda}^{QJ}(\omega) = 0  \,\,\,\,\,\,\,\,\,\,\,\,\,\,\,
(\omega_n \neq 0) .  
\label{fo}
\end{align}
Inserting here the RPA form (\ref{rpag2}) and using the pole behavior of
the reduced T-matrix (\ref{pole}), we obtain
\begin{align}
\sum_{\lambda} N_{\lambda}(\omega_n) \, \pi_{\lambda}^{QJ}(\omega_n) 
= 0  \,\,\,\,\,\,\,\,\,\,\,\,\,\,\,
(\omega_n \neq 0) . \label{fo1}
\end{align}
The physical meaning of this relation is that the total angular momentum operator
${\bold J} = {\bold J}_p + {\bold J}_n$ 
cannot cause transitions from the ground state to a state with finite excitation
energy. This follows easily from (\ref{fo1}) by noting that the transition matrix element of
the $K=1$ component of any tensor operator $W_{\tau}^1$ from the ground state
to an excited state is given by (see Fig.4)
\begin{align}
\langle \omega_n, K=1|{W}^1_{\tau}| 0 \rangle
= \frac{1}{\sqrt{2 \omega_n}} \, N_{\tau}(\omega_n) \, \pi^{QW}_{\tau}(\omega_n) \, .   
\label{trans}
\end{align}
For the case $W=J$ we obtain the physical interpretation of the relation
(\ref{fo1}) as explained above. 
\begin{figure}[tbp]
\begin{center}
\includegraphics[width=\columnwidth]{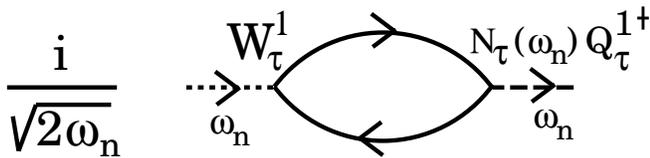}
\caption{Graphical representation of the transition matrix element, Eq.(\ref{trans}).}
\end{center}
\end{figure}
 
 
In Appendix A, 
the creation and annihilation operators for
the Goldstone modes and the resulting isoscalar part of the effective rotational Hamiltonian are presented
\footnote{In Appendix A, we also take the chance to correct some 
formal inaccuracies which occurred in Ref.\cite{BAERW} in connection with the 
Goldstone limit $\omega_0 \rightarrow 0$.}. 
In particular, it is shown that the total angular momentum operator determines the form of 
the creation and annihilation operators for the zero modes and the effective rotational Hamiltonian. 
In Appendix B, a simple analytic
approximation, which takes into account only the isoscalar zero modes and the 
isovector low energy modes, is used to present a similar discussion for the isovector
scissors mode.
In particular, it is shown that the isovector combination of $J_p$ and
$J_n$ determines the form of the rotational part of 
the creation and annihilation operators and the effective rotational Hamiltonian.
By adding the contributions from the Goldstone and scissors degrees
of freedom, we derive an effective Hamiltonian of the 2-rotor form. 

\section{M1 sum rules}
\setcounter{equation}{0}

In this Section we wish to discuss the inverse energy weighted (IEW) and
energy weighted (EW) magnetic sum rules as they emerge in the simple framework
discussed in the previous Sections. Our main interest is to see how the
presence of the spin part of the M1 operator changes the sum rules,
and to gain some intuitive understanding of the results.

We express the $K=1$ component of the M1 operator in the form
\begin{align}
M^1 = \sum_{\tau} \left(g_{\ell \tau} L^1_{\tau} + g_{s \tau} S^1_{\tau} \right)
\equiv \sum_{\tau} \left(g_{\tau} J^1_{\tau} + h_{\tau} S^1_{\tau} \right) \,,
\label{m1}
\end{align}
where the values of the free proton and neutron $g$-factors are
$g_{\ell p}^{\rm free} =1$, $\,\,g_{\ell n}^{\rm free} =0$, $\,\,g_{s p}^{\rm free} =5.58$,
$\,\,g_{s n}^{\rm free} =-3.82$. In the second equality of (\ref{m1}) we used 
$L^1_{\tau} = J^1_{\tau} - S^1_{\tau}$, and defined 
\begin{align}
g_{\tau} \equiv g_{\ell \tau}\,, \,\,\,\,\,\,\,\,\,\,
h_{\tau} = g_{s \tau} - g_{\ell \tau} \,.  \label{gh}
\end{align}
The free nucleon values of $h_{\tau}$ are therefore $h_{p}^{\rm free}=4.58$ and $h_n^{\rm free}=-3.82$.
We remark that in principle the $g$-factors appearing in (\ref{m1}) should be regarded as
effective ones\cite{MIG}, which include short range processes (tensor correlations,
meson exchange currents, etc), which are not taken into account by the RPA-type
correlations.  

To derive the sum rules, we follow Ref.\cite{BAERW} and consider the following 2-point 
function with external M1 operators:
\begin{align}   
\Pi^{\rm MM}(\omega) &=
i \, \int {\rm d} \tau \, e^{i \omega \tau} \langle 0 | 
T \left( M^{1 \dagger}(t') \, M^1(t) \right) |0 \rangle,
\nonumber \\
&= - 2 \sum_n \, |\langle n | M^1 | 0 \rangle |^2  
\, \frac{\Omega_n}{\omega^2 - \Omega_n^2 + i \delta}\, ,
\label{specm}
\end{align}
where $\tau = t'-t$, and we used the notations of Eq.(\ref{spec1}). 
The IEW and EW sum rules can then be expressed as follows\cite{LS}:
%
\begin{align}
S_{\rm IEW} &\equiv  2 \sum_{\Omega_n > 0} \frac{| \langle n| M^1 | 0\rangle |^2}{\Omega_n} 
= \lim_{\omega \rightarrow 0} \Pi^{MM}(\omega),
\label{iew}  \\
S_{\rm EW} &\equiv  2 \sum_{\Omega_n>0} | \langle n| M^1 | 0\rangle |^2 \, \Omega_n 
= - \lim_{\omega \rightarrow \infty} \omega^2 \, \Pi^{MM}(\omega) .
\label{ew}
\end{align}
The quantity $\Pi^{\rm MM}(\omega)$ of (\ref{specm}) is related to the correlators
$\Pi^{W' W}_{\tau \lambda}(\omega)$ of (\ref{corr}) by
\begin{eqnarray}
\Pi^{\rm MM}(\omega) &=& \sum_{\tau \lambda} 
\left( g_{\tau} \Pi^{JJ}_{\tau \lambda}
(\omega) g_{\lambda} + g_{\tau} \Pi^{JS}_{\tau \lambda} (\omega) h_{\lambda} \right.
\nonumber \\
& &\left. + h_{\tau} \Pi^{SJ}_{\tau \lambda}(\omega) g_{\lambda}
+ h_{\tau} \Pi^{SS}_{\tau \lambda}(\omega) h_{\lambda} \right)\,,   
\nonumber  \\ 
\label{pimm}
\end{eqnarray}
and therefore the calculation of M1 sum rules reduces to the calculation of the
correlators $\Pi^{R' R}_{\tau \lambda}(\omega)$ in the limits indicated by Eq.(\ref{iew})
and (\ref{ew}), where $R$ or $R'$ denotes $J$ or $S$. In our simple RPA framework, those
correlators are given by (\ref{rpag2}), i.e.,
\begin{align}
\Pi^{R' R}_{\tau \lambda}(\omega) = \delta_{\tau \lambda} \pi_{\tau}^{R' R}(\omega)
- \pi_{\tau}^{R' Q}(\omega) t_{\tau \lambda}(\omega) \pi_{\lambda}^{QR}(\omega) \,.
\label{rr}
\end{align}

\subsection{Inverse energy weighted sum rule}

For the IEW sum rule, we need 
the correlator (\ref{rr}) for $\omega=0$. Because of the relation (\ref{rel1}),
the bubble graphs in the second term on the r.h.s. of (\ref{rr}) bring in a
factor $\omega^2$, and therefore only the singular term in the reduced
T-matrix of Eq.(\ref{sing}) contributes. Using the relation (\ref{rgv}),
we obtain the very simple result
\begin{align}
\Pi^{R' R}_{\tau \lambda}(0) = \delta_{\tau \lambda} \pi_{\tau}^{R' R}(0)
- \frac{\pi_{\tau}^{R' J}(0) \pi_{\lambda}^{JR}(0)}{I^J} \,.
\label{simple}
\end{align}
By taking $R'$ or $R$ equal to $J$ in this expression, and using the definition of the
moments of inertia as given by Eq.(\ref{ingpn}) and (\ref{ingtot}), it is then evident that
\begin{align}
\sum_{\lambda} \Pi^{R' J}_{\tau \lambda}(0) =
\sum_{\tau} \Pi^{J R}_{\tau \lambda}(0) = 0 \,.   \label{conf}
\end{align}
This is a special case of the condition (\ref{ang}) for $\omega=0$, which follows
from angular momentum conservation. The relation (\ref{conf}) implies that only the 
isovector combination of the $J_{p}$ and $J_n$ contributes.
It is therefore useful to split the orbital $g$-factors into
isoscalar and isovector parts according to
\begin{align}
g_{\tau} = g_0 + t_{\tau} g_1 \,,  \label{iso}
\end{align}
where $t_p=1$ and $t_n=-1$. The factors $g_0$ and $g_1$ are the isoscalar and
isovector orbital $g$-factors, with the free nucleon values given by
$g_0^{\rm free}=g_1^{\rm free}=0.5$. Because of (\ref{conf}), we
can effectively replace $g_{\tau} \rightarrow t_{\tau} g_1$ in the correlator (\ref{pimm})
for $\omega=0$. 
On the other hand, the terms involving the spin operators give both isoscalar and isovector 
contributions, and the result does not simplify by splitting 
$h_{\tau}$ into isoscalar and isovector pieces
like in (\ref{iso}). We therefore express the final result by using the factors 
$g_1, h_p, h_n$ in the following way (see Appendix D): 
%
%
%
\begin{align}
&S_{\rm IEW} = \frac{4 I_p^J  I_n^J}{I^J} g_1 \left(
g_1 + h_p \frac{\hat{I_p^M}}{I_p^J} - h_n \frac{\hat{I_n^M}}{I_n^J} \right) 
\label{scsg} \\
&+ h_p^2 I_p^S + h_n^2 I_n^S - \frac{|h_p I_p^M + h_n I_n^M|^2}{I^J} \,.
\label{ss}
\end{align}  
The term $\propto g_1^2$ in (\ref{scsg}) is the familiar scissors IEW sum rule \cite{SRM,BAERW}.
In order to express the contributions from the spin part of the M1 operator, 
we defined the spin and mixed moments of inertia ($I^S_{\tau}$ and $I^M_{\tau}$) as follows:  
\begin{align}
I^S_{\tau} \equiv \pi_{\tau}^{SS}(0) &= 2 \sum_{(\alpha i) \in \tau} \frac{
|\langle \alpha | S^1_{\tau} | i \rangle|^2} {\omega_{\alpha i}} 
\label{ingspn} \\
I^M_{\tau} \equiv \pi_{\tau}^{JS}(0) &= 2 \sum_{(\alpha i) \in \tau}
\langle \alpha | J^1_{\tau} | i \rangle^* \langle \alpha | S^1_{\tau} | i \rangle  
\frac{1}{\omega_{\alpha i}} 
\label{ingmpn} \\
&= \pi_{\tau}^{LS}(0) + \pi_{\tau}^{SS}(0) \,, 
\label{splitb}
\end{align}
and $\hat{I_{\tau}^M}$ in (\ref{scsg}) means the real part
\footnote{The phase conventions used in this paper are the same as in Ref.\cite{BM},
and $I_{\tau}^M$ is actually real. Formally this is seen by noting that the
expansion coefficients of deformed states in terms of spherical states 
(see Eq.(\ref{expand})) are real, and matrix elements of the spherical components of
angular momentum operators between the spherical states are also real.} 
of $I_{\tau}^M$.

In Appendix B we show that the contribution (\ref{scsg}), which arises from the terms 
$JJ$, $JS$ and $SJ$ in the correlator
(\ref{pimm}), can be reproduced in a simple analytic approximation which takes into account 
only the low energy scissors mode, in addition to the Goldstone mode. This makes it
plausible that the contribution (\ref{scsg}) arises mainly
from the low energy scissors mode in the spectral sum (\ref{iew}). 
The term (\ref{ss}), on the other hand, cannot be reproduced in this way.
In fact, because this term arises from the $SS$ term in the M1 correlator
(\ref{pimm}), it will be modified by spin dependent interactions, and cannot
be reliably discussed within our ``minimal model''.
Nevertheless, it is worth while to point out that the $QQ$-type interaction, which
leads to the last term of (\ref{ss}) in the RPA, gives a negative contribution to
the spin-spin part of the IEW sum rule, that is, the spin part of the sum rule is smaller
than the naive (non-interacting) part, which is given by the first two terms in
(\ref{ss})\footnote{The sum of the 3 terms in (\ref{ss}) is of course positive,
because it corresponds to the sum rule (\ref{iew}) with $M^1$ replaced by 
${\displaystyle \sum_{\tau} h_{\tau} S_{\tau}^1}$.}.    
 


If the spin-orbit interaction is assumed to be zero, 
all spin correction terms in (\ref{scsg}), and also all terms in (\ref{ss}), vanish. 
This is easily seen because in this
case the relation $\left[H_0, S_{\tau}^1\right]=0$ is satisfied, which implies 
$\langle \alpha | S_{\tau}^1 | i \rangle = 0$ for $\omega_{\alpha i} >0$. That is, 
if $S_{\tau}^1$ commutes with the mean field Hamiltonian, it cannot cause
particle-hole excitations with finite excitation energy. The
spectral representation (\ref{pi}) then indicates that $\pi_{\tau}^{R' S}(\omega)=0$, where
$R'=J$ or $S$
\footnote{Zero energy excitations ($\omega_{\alpha i}=0$) do not contribute from the outset,
because the correlator (\ref{pi}) involves an explicit factor ${\omega}_{\alpha i}$ for
external operators with the same $T$-symmetry. If one needs the limit $\omega\rightarrow 0$,
this factor should be taken to zero first. 
(For the same reason, the Goldstone
modes do not contribute to the correlator (\ref{pimm}).)
We also note that the argument based on the commutation relation given above can be applied 
also to the interacting case,
where ${\bold J}$ is the conserved quantity. The resulting identity was already
expressed by Eq.(\ref{ang}).}.   
Therefore, if the spin-orbit interaction is set to zero, the IEW sum rule
reduces to the pure orbital result\cite{BAERW}, as expected.
We also remark that, for the case of spherical symmetry, 
all terms except the pure spin terms (the first 2 terms in (\ref{ss}))   
vanish, because in this case
the relation $\left[H_0, J_{\tau}^1 \right]=0$ is satisfied, and a similar
argument as given above implies that all bubble graphs which involve at least
one operator $J_{\tau}^1$ vanish. In particular, the mixed moments of inerita,
which are responsible for the spin correction terms in
in (\ref{scsg}), separately vanish in the limit of spherical symmetry.   

From the above discussions it is clear that the correction terms in (\ref{scsg}), which mainly
arise from the coupling of the external spin operator to the scissors mode, are non-zero 
only in the case of finite spin-orbit interaction and finite deformation. 
This is also clear from the basic formulae for the matrix elements of the
orbital and spin angular momentum operators between deformed single particle states, which
are collected in Appendix C. 
The sign of the mixed polarization 
$\hat{I}_{\tau}^M$ is, however, a delicate matter, because it depends on 
the quantum numbers of the valence nucleon states, the degree of deformation,
and also further model assumptions on the mean field Hamiltonian. 
If the sign of ${\hat I}_{\tau}^M$ 
turns out to be negative, the spin correction terms in (\ref{scsg}) will decrease the value 
of $S_{\rm IEW}$, which would help to explain the discrepancy between the theoretical and
experimental values. In order to see the sign and the size of the
correction terms, a numerical analysis in a Nilsson-type approach\cite{NI} is necessary, 
which, however, goes beyond the purpose of the analytical approach pursued in this paper.

\subsection{Energy weighted sum rule}

The evaluation of the EW sum rule (\ref{ew}) is again based on the
expressions (\ref{pimm}) and (\ref{rr}) for the correlators. Because now we have
to consider the limit $\omega \rightarrow \infty$, the $T$-matrix $t_{\tau \lambda}$
in the RPA correlators (\ref{rr}) reduces to the 4-Fermi coupling constant 
$\chi_{\tau \lambda}$, as is clear from the expression (\ref{t1}). By using the
expression (\ref{pi}) for the bubble graphs, we obtain
\begin{align}
& - \lim_{\omega \rightarrow \infty} \omega^2 \Pi_{\tau \lambda} ^{R' R}(\omega)
= 2 \delta_{\tau \lambda} \sum_{\alpha i \in \tau}
\langle \alpha | R'^1 |i \rangle ^*  \langle \alpha | R^1 |i \rangle \, \omega_{\alpha i}
\label{ewfree} \\
&+ 4 \sum_{\alpha i \in \tau} \sum_{\alpha' i' \in \lambda} 
\langle \alpha | R'^1 |i \rangle ^*  \langle \alpha | Q^1 |i \rangle
\, \chi_{\tau \lambda}   \langle \alpha' | Q^1 |i' \rangle ^* 
\langle \alpha' | R^1 |i' \rangle \,,  \label{ewint} 
\end{align}
where $R$ (or $R'$) denotes either $J$ or $S$. For the case $R'=R=J$, one can use the
relation (\ref{rel}) to replace the operators $J_{\tau}$ by the quadrupole operators
$Q_{\tau}$, and
make use of the self consistency relation (\ref{gap}) to obtain the result derived
in Ref.\cite{BAERW}. A similar calculation is   
possible also for the case where $R$ (or $R'$) is the spin operator $S$. The
derivation is explained in Appendix D, which in particular shows that there are
no crossing terms between $J$ and $S$ in the EW sum rule, i.e.,
\begin{align}
\lim_{\omega \rightarrow \infty} \omega^2 \Pi_{\tau \lambda} ^{J S}(\omega) = 0 \,.
\label{zer}
\end{align}
The final result becomes
\begin{align}
S_{\rm EW} = -12 \, g_1^2 \, \langle Q_p^0 \rangle \chi_{pn} \langle Q_n^0 \rangle
+ \frac{2}{3} \sum_{\tau} h_{\tau}^2 \gamma_{\tau} \langle {\bold L}_{\tau} \cdot
{\bold S}_{\tau} \rangle\,.
\label{ewfin}
\end{align}
Here $\gamma_{\tau}$ is the strength of the spin-orbit interaction, see Eq.(\ref{h0}).

The second term in (\ref{ewfin}), which comes from the spin part of the M1
operator, is positive because $\langle {\bold L}_{\tau} \cdot
{\bold S}_{\tau} \rangle > 0$. (A formal argument for this is given in Appendix D.)
However, as we noted already in connection with the IEW sum rule, 
spin dependent interactions will give additional contributions to
the $S-S$ correlator, which will modify the result (\ref{ewfin}). 
Unlike the case of the 
IEW sum rule, there is no spin effect in the EW sum rule   
which reflects the excitation of the low energy scissors mode.

\section{Summary}

In this paper we used a simple field theory model for nucleons to get analytic insight
into the role of the spin for low energy nuclear rotational states.
Our starting point was a nuclear Hamiltonian which explicitly breaks the
separate spin and orbital symmetries due to the presence of the spin-orbit
potential from the outset, and we used the mechanism of spontaneous symmetry
breaking to generate a deformed mean field.
The residual interaction, for which we used a simple $QQ$ interaction, 
was taken into account in the framework of the RPA.
Our main focus was on the inverse energy weighted (IEW) magnetic sum rule, for which
we found that the spin part of the M1 operator gives nonzero contributions,
even in the absence of spin dependent residual interactions.
We argued that the spin-orbit potential gives rise to new 
non-interacting polarizations $\pi^{R' R}$, where one of the operators $R'$, $R$ is a spin
operator. In particular, the polarizations $\pi^{JS}$ modify
that part of the IEW sum rule which arises mainly from the excitation
of the low energy scissors mode. 
In order to assess the effect of those spin corrections quantitatively, however,  
a numerical approach is necessary. 

By using a simple analytic approximation, we also
confirmed that the structures of the isoscalar Goldstone and isovector
scissors modes, and of the resulting effective rotational Hamiltonian, are 
similar to the case of the pure orbital description, if one replaces the
orbital angular momenta by the total angular momenta. 
 
Finally we wish to address the question whether it is possible to 
construct models where the pure orbital nature of the low energy scissors
mode is maintained. This is indeed possible if one starts from a Hamiltonian which conserves
the orbital and spin symmetries separately, and generates the spin-orbit
interaction by the mechanism of spontaneous symmetry breaking. The order parameter
which describes this breaking is the ground state expectation value
$\langle {\bold L} \cdot {\bold S}\rangle $, and the three associated Goldstone bosons 
are represented by ${\bold L} \times {\bold S}$. If combined with the pattern of dynamical
symmetry breaking used in the present paper, the associated 5 isoscalar Goldstone
degrees of freedom lead to separate 
orbital and spin rotational bands, and for each there exists an isovector counterpart
at low excitation energy
\footnote{A condensate of the above kind
in the particle-particle channel ($^3P_0$ condensate)  
has been used for the description of superfluid
Helium 3 in the B phase\cite{NA}.  The spin rotational bands mentioned above might be 
connected to the so called magnetic
rotational bands observed in weakly deformed nuclei\cite{MROT}.}.
We plan to present such an approach in a separate paper\cite{TWO}.

\vspace{1 cm}

\noindent
{\sc Acknowledgements}

One of the authors (W.B.) wishes to express his thanks to Prof. K. Yazaki for 
very helpful discussions.

\appendix

\section{Creation operators for the Goldstone modes and isoscalar effective 
rotational Hamiltonian}
\setcounter{equation}{0}

We first recall that the $T$-symmetry relation for the particle-hole matrix elements 
of any operator 
$A$ is given by (see footnote 1 of Sect. II)
\begin{align}
\langle \overline{i}|{A}|\overline{\alpha} \rangle = t_{A} \langle \alpha | A | i \rangle \,, 
\label{ts}
\end{align}
where the time-reversed single particle states $(\overline{\alpha} \overline{i})$ have the
opposite values of $j_z$ but the same energies as $(\alpha i)$ for the case of axial
symmetry, and $t_A = \pm 1$. For the case of the $T$-even quadrupole operator we have $t_Q = +1$.
  
In order to establish the connection of the BS formalism used in the main text to the more conventional 
RPA formalism\cite{RO,RS},
we return to the expression (\ref{trans}) for the transition matrix element of a tensor operator
$W_{\tau}^1$ and insert the spectral 
representation of the correlator $\pi_{\tau}^{QW}$ in the form 
\begin{eqnarray}
\pi_{\tau}^{QW}(\omega) &=& - \sum_{(\alpha i) \in \tau} \left[
\frac{\langle \alpha|W^1|i \rangle \langle \alpha|Q^1|i \rangle^*}
{\omega - \omega_{\alpha i} + i \delta} \right.
\nonumber \\ 
&-&  \left.\frac{\langle i|W^1|\alpha \rangle \langle i|Q^1|\alpha \rangle^*}
{\omega + \omega_{\alpha i} - i \delta} \right] \,.
\label{corrf}
\end{eqnarray}
By using the $T$-symmetry relation (\ref{ts}) for the second term in (\ref{corrf}) to combine it with the
first one, it is easy to see that this relation is the
same as (\ref{pi}) of the main text.  
Using (\ref{corrf}), the transition matrix element (\ref{trans}) then takes the form
\begin{multline}
\langle \omega_n, K=1 | W_{\tau}^1 | 0 \rangle = \\
\sum_{(\alpha i) \in \tau}
\left[ \left(Y_{\tau}^*\right)_{\alpha i}(\omega_n) \left(W_{\tau}^1\right)_{\alpha i}
+ \left(Z_{\tau}^*\right)_{\alpha i}(\omega_n) \left(W_{\tau}^1\right)_{i \alpha} \right]\,.
\label{tr}
\end{multline}
We used the notation $\left( A \right)_{\alpha i} = \langle \alpha| A |i \rangle$
for the particle-hole matrix elements of an operator $A$, and defined the $K=1$
components of the RPA amplitudes by
\begin{align}
\left(Y_{\tau}\right)_{\alpha i}(\omega_n) &= \frac{-1}{\sqrt{2 \omega_n}}
\frac{N_{\tau}(\omega_n) \left(Q_{\tau}^1\right)_{\alpha i}}{\omega_n - \omega_{\alpha i}
+ i \delta} \,, \label{yrpa} \\ 
\left(Z_{\tau}\right)_{\alpha i}(\omega_n) &= \frac{1}{\sqrt{2 \omega_n}}
\frac{N_{\tau}(\omega_n) \left(Q_{\tau}^1\right)_{i \alpha}}{\omega_n + \omega_{\alpha i}
- i \delta} \,. \label{zrpa} 
\end{align}
In the conventional RPA formalism, the states $|\omega_n, K=1 \rangle$ are then expressed by
\begin{align}
|\omega_n, K=1 \rangle = {\cal O}^{\dagger}(\omega_n, K=1) |0 \rangle \,,
\label{state}
\end{align}
where ${\cal O}^{\dagger} = \sum_{\tau} {\cal O}_{\tau}^{\dagger}$ with
\begin{align}
{\cal O}^{\dagger}_{\tau}(\omega_n, K=1) = \sum_{\alpha i}
\left[\left(Y_{\tau}\right)_{\alpha i}(\omega_n) B^{\dagger}_{\alpha i} - 
\left(Z_{\tau}\right)_{\alpha i}(\omega_n) B_{\alpha i} \right]\,.
\label{cr}
\end{align}
Here we introduced the creation and annihilation operators for a particle-hole pair by
\begin{align}
B^{\dagger}_{\alpha i} = a^{\dagger}_{\alpha} a_i \,,
\,\,\,\,\,\,\,\,\,\,B_{\alpha i} = a^{\dagger}_i a_{\alpha} \,.
\end{align}

The RPA consists in 
\begin{align}
\left[ B_{\alpha i} , B^{\dagger}_{\beta j} \right] \stackrel{{\rm RPA}}{\equiv} 
\delta_{\alpha \beta} \delta_{i j} \,,
\label{comm}
\end{align}
while the commutators of two creation (or two annihilation) operators vanish.
If we also expand the external operator $W_{\tau}^1$ in terms of the creation and
annihilation operators as
\begin{align}
W_{\tau}^1 = \sum_{(\alpha i) \in \tau} \left[ \left( W^1\right)_{\alpha i} B^{\dagger}_{\alpha i}
+ \left( W^1\right)_{i \alpha} B_{\alpha i}  \right] \,
\label{expandw}
\end{align}
and require the RPA (\ref{comm}), it is easy to check that 
\begin{align}
\langle \omega_n, K=1 | W_{\tau}^1 | 0 \rangle = \langle 0| \left[ {\cal O}_{\tau}(\omega_n, K=1) , 
W_{\tau}^1 \right] |0 \rangle  
\label{tr1}
\end{align} 
gives the same result as (\ref{tr}). 

By using the forms (\ref{yrpa}) and (\ref{zrpa}) and the $T$-symmetry
relation (\ref{ts}), it is also easy to confirm the familiar normalization relation for the
RPA amplitudes:
\begin{multline}
\sum_{\tau} \sum_{\alpha i} \left( |\left(Y_{\tau}\right)_{\alpha i}(\omega_n)|^2
- |\left(Z_{\tau}\right)_{\alpha i}(\omega_n)|^2 \right) =  \\
= \sum_{\tau} N_{\tau}^2(\omega_n) \pi_{\tau}^{QQ}{}'(\omega_n) = 1\,,
\label{conf1}
\end{multline}
where the derivatives of the $QQ$ bubble graphs were given in (\ref{pip}),
and in the last step of (\ref{conf1}) we used the condition (\ref{norm}) of the main text.

In the rest of this Appendix, we concentrate on the case of the $K=\pm 1$ Goldstone modes ($n=0$),
where the limit $\omega_0 \rightarrow 0$ needs special care. We first wish to confirm that our method gives
creation and annihilation operators for those modes, which are independent of each other and 
satisfy the correct boson commutation relations. By using the relation (\ref{rel}) in the
expressions (\ref{yrpa}) and (\ref{zrpa}) of the RPA amplitudes, and the form (\ref{rgv}) of
the normalization factors, we obtain the following expressions up to the order $\sqrt{\omega_0}$:
\begin{align}
\left( Y_{\tau}\right)_{\alpha i}(\omega_0) &= \frac{-1}{\sqrt{2 \omega_0 I^J}}
\left(1 + \frac{\omega_0}{\omega_{\alpha i}} \right) \langle \alpha | J_{\tau}^1 | i \rangle\,,        
\label{y0} \\ 
\left( Z_{\tau}\right)_{\alpha i}(\omega_0) &= \frac{1}{\sqrt{2 \omega_0 I^J}}
\left(1 - \frac{\omega_0}{\omega_{\alpha i}} \right) \langle i | J_{\tau}^1 | \alpha \rangle\,.         
\label{z0} 
\end{align}
The important points to note for the subsequent calculations are: (i) The squares of those amplitudes
contain, besides the divergent terms $\propto 1/\omega_0$, also finite terms of order $1$. 
(ii) In the subsequent expressions, the divergent terms cancel between the forward and backward 
amplitudes, on account of the $T$-symmetry relation (\ref{ts}) for the $T$-odd operators $J_{\tau}^K$. 
By noting these two points, it is then easy to confirm that (\ref{y0}) and (\ref{z0}) and
their $K=-1$ counterparts , which are obtained by $J_{\tau}^1 \rightarrow - J_{\tau}^{-1}$, 
satisfy the following relations: \\

\noindent
(1) Normalization of RPA amplitudes: 
\begin{multline}
\sum_{\tau} \sum_{\alpha i} \left( |\left(Y_{\tau}\right)_{\alpha i}(\omega_0)|^2
- |\left(Z_{\tau}\right)_{\alpha i}(\omega_0)|^2 \right) =  \\
\hspace{-2cm} = \frac{2}{I^J} \sum_{\tau} \sum_{(\alpha i)\in \tau} \frac{1}{\omega_{\alpha i}} 
|\langle \alpha|J_{\tau}^1|i \rangle|^2 = 1\,,
\label{c1}
\end{multline}
where in the last step we used the form (\ref{ingtot}) of the total moment of inertia.
The same argument can be applied also to the $K=-1$ mode.\\

\noindent
(2) Orthogonality of $K=1$ and $K=-1$ modes: \\

The RPA amplitudes for the $K=-1$ Goldstone mode are obtained by replacing 
$J_{\tau}^1 \rightarrow - J_{\tau}^{-1}$
in the expressions (\ref{y0}) and (\ref{z0}). Indicating the $K$-values explicitly, we obtain
\begin{multline}
\sum_{\tau} \sum_{(\alpha i)\in \tau} \left[
\left(Y_{\tau}^*\right)_{\alpha i}(\omega_0, K=-1) \left(Y_{\tau}\right)_{\alpha i}(\omega_0, K=1) \right. \\
\left. - \left(Z_{\tau}^*\right)_{\alpha i}(\omega_0, K=-1) \left(Z_{\tau}\right)_{\alpha i}(\omega_0, K=1) \right]   
\\
= \frac{-2}{I^J} \sum_{\tau} \sum_{(\alpha i)\in \tau} \frac{1}{\omega_{\alpha i}}
\langle \alpha|J_{\tau}^{-1}|i \rangle^* \langle \alpha|J_{\tau}^{1}|i \rangle = 0 \,,
\label{c2}
\end{multline}
where in the last step we used the fact that for axial symmetry $j^z \equiv m$ of the
single particle states is a good quantum number. This implies that the state 
$|\alpha \rangle$ cannot have the values
$m_{i}+1$ and $m_{i}-1$ at the same time.\\

\noindent
(3) Boson commutation relations for the Goldstone modes: \\

The creation and annihilation operators for the $K=1$ Goldstone mode are obtained by inserting the
forms (\ref{y0}) and (\ref{z0}) into (\ref{cr}) and the h.c. of (\ref{cr}). This gives
\begin{eqnarray}
\lefteqn{{\cal O}^{\dagger}(\omega_0, K=1) = \frac{-1}{\sqrt{2 \omega_0 I^J}} \sum_{\tau}
\sum_{(\alpha i) \in \tau}}  \nonumber \\
& & \hspace{-1cm} \times \left[ \left(1 + \frac{\omega_0}{\omega_{\alpha i}} \right) 
\langle \alpha|J_{\tau}^1 |i \rangle
B^{\dagger}_{\alpha i} + \left(1 - \frac{\omega_0}{\omega_{\alpha i}} \right) 
\langle i|J_{\tau}^1 |\alpha \rangle B_{\alpha i} \right]  \nonumber \\ 
\label{o0d} \\ 
\lefteqn{{\cal O}(\omega_0, K=1) = \frac{-1}{\sqrt{2 \omega_0 I^J}} \sum_{\tau}
\sum_{(\alpha i) \in \tau}}  \nonumber \\
& & \hspace{-1cm} \times \left[ \left(1 + \frac{\omega_0}{\omega_{\alpha i}} \right) \langle \alpha|J_{\tau}^1 
|i \rangle^* B_{\alpha i} + \left(1 - \frac{\omega_0}{\omega_{\alpha i}} \right) 
\langle i|J_{\tau}^1 |\alpha \rangle^* B^{\dagger}_{\alpha i} \right]\,.  \nonumber \\ 
\label{o0}
\end{eqnarray}
The operators for the $K=-1$ mode are obtained by $J_{\tau}^1 \rightarrow - J_{\tau}^{-1}$ in 
the above expressions. 
By using the $T$-symmetry relation (\ref{ts}) for the operators $J_{\tau}$, it is then easy 
to confirm that
\begin{eqnarray}
\lefteqn{\left[ {\cal O}(\omega_0,K=1), {\cal O}^{\dagger}(\omega_0,K=1) \right]} 
\nonumber \\
& & \hspace{-1cm} 
= \frac{2}{I^J} \sum_{\tau} \sum_{(\alpha i)\in \tau} \frac{1}{\omega_{\alpha i}}
|\langle \alpha | J_{\tau}^1 | i \rangle |^2 = 1 \,,
\label{commg}
\end{eqnarray}
where in the last step we used the form (\ref{ingtot}) of the total moment of inertia.
Similar arguments hold also for the case $K=-1$. The conservation of $j^z$ of the single
particle states can be used to show that the commutators between the operators for 
the $K=1$ and $K=-1$ modes vanish, similar to the argument given below Eq.(\ref{c2}). 
The commutation relation (\ref{commg})
can be used to confirm the correct normalization of the Goldstone boson states
$|\omega_0, K \rangle = {\cal O}^{\dagger}(\omega_0, K) |0 \rangle$, namely
$\langle \omega_0, K | \omega_0, K \rangle = 1$.\\


\noindent
Finally in this Appendix, we note that the Goldstone modes give the following
contribution to the effective rotational Hamiltonian of the system:
\begin{align}
H_{\rm rot}(\omega_0) 
&= \omega_0 \sum_{K=\pm 1} {\cal O}^{\dagger}(\omega_0, K) {\cal O}(\omega_0, K) 
\label{hr1} \\
&= \frac{\left(J^x\right)^2 + \left(J^y\right)^2}{2 I^J}
= \frac{\vec{J}^2 - \left(J^z\right)^2}{2 I^J} \,,  
\label{hr2}
\end{align}
where we used the forms (\ref{o0d}) and (\ref{o0}). We see that, because of the
overall factor $\omega_0$, only the singular terms in (\ref{o0d}) and (\ref{o0})
contribute to the effective rotational Hamiltonian, although it would be
meaningless to take the
limit $\omega_0 \rightarrow 0$ directly in those expressions.


\section{Scissors mode and M1 sum rule}
\setcounter{equation}{0}

In this Appendix we discuss a simple analytic approximation for the isovector scissors mode.
One of our intentions is to show that this approximation can reproduce the contribution 
(\ref{scsg}) to the IEW 
M1 sum rule, which suggests that also in the ``exact'' RPA approach of the main text this
term comes mainly from the scissors mode contribution to the transition matrix element depicted
in Fig. 4.

Analytic approximations are obtained by assuming simple pole forms for the bubble 
graphs (\ref{pi})\cite{BAERW}. 
Here we discuss the simplest approximation of a 
one-pole form, where the energies of the particle-hole states 
are replaced by an average excitation energy:
$\omega_{\alpha i} \rightarrow e_0$. In this approximation, which is similar to the
``closure approximation'' (effective energy denominators) used in different 
contexts\cite{CL}, the pole equation 
(\ref{det}) gives
only two solutions, which correspond to the isoscalar Goldstone mode 
and the low energy isovector scissors mode.  

\subsection{Excitation energy and vertex functions}

To obtain the excitation energy and the vertex functions for the scissors mode
in the schematic model, we need the form of $\pi_{\tau}^{QQ}$,
see (\ref{pi}). If we use the identity (\ref{rel}) and make the one-pole approximation as
explained above, it can be expressed
in terms of the moments of inertia (\ref{ingpn}):
\begin{align}
\pi_{\tau}^{QQ}(\omega) = \frac{-1}{3 \beta_{\tau}^2} \frac{e_{0\tau}^4}
{\omega^2 - e_{0 \tau}^2} I_{\tau}^J   \label{pi0} 
\end{align}
Inserting this form into the pole equation (\ref{det}) we obtain
\begin{multline}
\left(\omega^2 - e_{0p}^2 \left(1 + \frac{\chi_{pp} e_{0p}^2 I^J_p}{3 \beta_p^2} \right) 
\right)  \\
\times \left(\omega^2 - e_{0n}^2 \left(1 + \frac{\chi_{nn} e_{0n}^2 I^J_n}{3 \beta_n^2} \right) 
\right) 
= \frac{\chi_{pn}^2 e_{0p}^4 e_{0n}^4 I^J_p I^J_n}{9 \beta_p^2 \beta_n^2}\,.  
\label{pole0}
\end{multline}
On account of the self consistency relation (\ref{gap1}) for exact symmetry 
($\varepsilon_{\tau}=0$), 
the Goldstone pole $\omega^2=0$ is a solution of (\ref{pole0}).
The other solution is then obtained as
\begin{align}
\omega_1^2 &= e_{0p}^2 \left(1 + \frac{\chi_{pp} e_{0p}^2 I^J_p}{3 \beta_p^2} \right) +  
e_{0n}^2 \left(1 + \frac{\chi_{nn} e_{0n}^2 I^J_n}{3 \beta_n^2} \right)  \nonumber \\
&= - \frac{\chi_{pn}}{3} \frac{e_{0p}^2 e_{0 n}^2}{\beta_p \beta_n} I^J \,,
\label{sol1}
\end{align}
where in the second step we used again the self consistency relation
(\ref{gap1}) for exact symmetry ($\varepsilon_{\tau}=0$).

By using the above relations, one can obtain the reduced vertex functions from (\ref{hombs}) and
(\ref{norm}). The result for the ratio follows from (\ref{hombs}) as
\begin{align}
\frac{N_p(\omega_1)}{N_n(\omega_1)} = - \frac{\omega_1^2 - e_{0n}^2}{\omega_1^2 - e_{0p}^2}
\frac{I_p^J e_{0p}^2 \beta_n}{I_n^J e_{0n}^2 \beta_p} \,,   
\label{ratn}
\end{align}
and the individual factors are then obtained from (\ref{norm}) as
\begin{align}
N_p(\omega_1) &= - \sqrt{\frac{3}{I^J}} \, \beta_p \, \sqrt{\frac{I^J_n}{I^J_p}} \, 
\frac{\omega_1^2 - e_{0p}^2}{e_{0 p}^2},  \label{normp} \\ 
N_n(\omega_1) &= \sqrt{\frac{3}{I^J}} \, \beta_n \, \sqrt{\frac{I^J_p}{I^J_n}} \,  
\frac{\omega_1^2 - e_{0n}^2}{e_{0 n}^2}\,.  \label{normn} 
\end{align}
The opposite sign for protons and neutrons indicates the isovector character of this 
mode. 

\subsection{Scissors contribution to the inverse energy weighted M1 sum rule}

The contribution of the scissors mode
$|\omega_1, K=1 \rangle$ to the IEW sum rule (\ref{iew}) is
given by
\begin{align}
S_{\rm IEW}^{\rm (sc)} = \frac{2}{\omega_1} \langle \omega_1, K=1 | 
\sum_{\tau} \left(g_{\tau} J_{\tau}^1 + h_{\tau} S_{\tau}^1\right) | 0 \rangle 
\nonumber  \\
\times \langle \omega_1, K=1 | 
\sum_{\tau} \left(g_{\tau} J_{\tau}^1 + h_{\tau} S_{\tau}^1 \right) | 0 \rangle^* \,,
\label{iew0}
\end{align}
where we used the second form given in (\ref{m1}) for the M1 operator.
To evaluate this, we need the form of the transition matrix element for an angular momentum operator
$R = J$ or $S$, which is obtained from (\ref{trans}) as
\begin{align}
\langle \omega_1, K=1 | R_{\tau}^1 | 0 \rangle = \frac{1}{\sqrt{2 \omega_1}}
\pi_{\tau}^{QR}(\omega_1) N_{\tau}(\omega_1)  \,.
\label{trans01}  
\end{align}
If we use the results (\ref{normp}) and (\ref{normn}) for the normalization factors,
and the following form of the polarization $\pi_{\tau}^{QR}$ in the one-pole approximation
\begin{align}
\pi_{\tau}^{QR}(\omega_1) = - \frac{\omega_1}{\sqrt{3} \beta_{\tau}} 
\frac{e_{0 \tau}^2}{\omega_1^2 - e_{0 \tau}^2} \pi_{\tau}^{JR}(0) \,,   
\label{piqr}
\end{align}
we obtain for the transition matrix elements 
\begin{align}
\langle \omega_1, K=1 | R_{p}^1 | 0 \rangle &=  \sqrt{\frac{\omega_1}{2 I^J}}
\sqrt{\frac{I_n^J}{I_p^J}} \, \pi_p^{JR}(0)  \label{tp} \\
\langle \omega_1, K=1 | R_{n}^1 | 0 \rangle &= - \sqrt{\frac{\omega_1}{2 I^J}}
\sqrt{\frac{I_p^J}{I_n^J}} \, \pi_n^{JR}(0)\,.  \label{tn}
\end{align}
The ordinary moments of inertia $I_{\tau}^J$ and the mixed ones $I_{\tau}^M$ have been defined in 
the main text as (see Eq.(\ref{ingpn}) and (\ref{ingmpn})) 
\begin{align}
I_{\tau}^J = \pi_{\tau}^{JJ}(0) \,, \,\,\,\,\,\,\,\,\,\,\,\,\,\,\,
I_{\tau}^M = \pi_{\tau}^{JS}(0) \,.  
\label{rept}
\end{align}
It is then clear from (\ref{tp}) and (\ref{tn}) that \\
$\langle \omega_1, K=1 | J_{p}^1 + J_n^1 | 0 \rangle=0$,
which is a consequence of angular momentum conservation. (See Eq.(\ref{fo1}) and the discussions below
that equation.) Therefore the isoscalar part of the first term 
in the M1 operator (\ref{m1}) does not contribute to the transition
matrix element, and we can replace $g_{\tau} \rightarrow t_{\tau} g_1$ in (\ref{iew0}), 
see Eq.(\ref{iso}).
The transition matrix element of the isovector $J$ - part is then obtained from (\ref{tp}) and
(\ref{tn}) as 
\begin{align}
\langle \omega_1, K=1 | \sum_{\tau}g_1 t_{\tau} J^1_{\tau} | 0 \rangle =
2 g_1 \sqrt{\frac{\omega_1}{2 J^J}} \sqrt{I_p^J \, I_n^J} \,.
\nonumber \\
\label{jv}
\end{align}
We insert this expression, as well as (\ref{tp}), (\ref{tn}) for the spin operator ($R=S$),
where also the isoscalar contribution is nonzero, into (\ref{iew0}), and obtain finally
\begin{align}
&S_{\rm IEW}^{\rm (sc)} = \frac{4 I_p^J  I_n^J}{I^J} g_1 \left(
g_1 + h_p \frac{\hat{I_p^M}}{I_p^J} - h_n \frac{\hat{I_n^M}}{I_n^J} \right) 
\label{scsg0} \\
&+ \frac{1}{I^J} | h_p \sqrt{\frac{I_n^J}{I_p^J}} I_p^M -  
h_n \sqrt{\frac{I_p^J}{I_n^J}} I_n^M|^2 \,.
\label{ss0}
\end{align}
If we compare this result to the full RPA result of our model, which was given by (\ref{scsg}) and (\ref{ss}), 
we see that the terms (\ref{scsg}) and (\ref{scsg0}), which originate from the products 
$J-J$, $J-S$ and $S-J$
in the sum rule, are identical, while the terms (\ref{ss}) and (\ref{ss0}), which come from
the product $S - S$, are different. This observation suggests that the
contribution (\ref{scsg}) of the full RPA sum rule comes mainly from the excitation of the
scissors mode, while the spin contribution (\ref{ss}) receives also contributions from 
other modes. 

\subsection{Creation operators for the scissors mode and two-rotor model}

Using the results of Sect. (B.1) for the excitation energy and the reduced vertex functions,
it is straight forward to determine the RPA amplitudes and the creation operators for the scissors
mode from (\ref{yrpa}), (\ref{zrpa}) and (\ref{cr}), and also the contribution to the
effective Hamiltonian of the system.  The calculation
is the same as in Ref.\cite{BAERW}, with the only difference that we now have the total angular 
momentum operators instead of the orbital ones. 
We therefore simply quote the results here.
 
%
The creation operator for the $K=1$ scissors mode 
is obtained from an expression similar to Eq.(\ref{cr}), with the RPA amplitudes
(\ref{yrpa}), (\ref{zrpa}) determined from the scissors vertex functions (\ref{normp}) and (\ref{normn}). 
It has the form ${\cal O}^{\dagger}(\omega_1, K=1) = 
{\cal O}^{\dagger}_J(\omega_1, K=1) + {\cal O}^{\dagger}_Q(\omega_1, K=1)$, where
\begin{align}   
&{\cal O}_J^{\dagger}(\omega_1, K=1) = 
\frac{-1}{\sqrt{2 \omega_1 I^J_{IV}}} 
\left( \frac{2 I^J_n}{I^J} J_p^1 - \frac{2 I^J_p}{I^J} J_n^1 \right), 
\label{oj} \\  
&{\cal O}_Q^{\dagger}(\omega_1, K=1) \no \\
%
&= \frac{1}{\sqrt{2 \omega_1 I^J_{IV}}} \sqrt{\frac{- \chi_{pn} I^J}{\beta_p \beta_n}}
\left( \frac{2 I^J_n}{I^J} \beta_n Q_p^1 - \frac{2 I^J_p}{I^J} \beta_p Q_n^1 \right). 
\label{oq}
\end{align}
%
Here the isovector moment of inertia is defined as
\begin{align}
I^J_{\rm IV} = \frac{4 I_p^J I_n^J}{I^J}\,.  
\label{iiv}
\end{align}
The form (\ref{oj}) of the creation operator clearly indicates the scissors character of the mode. 
It is important, however, to note that it involves the total angular momenta ${\bold J}_{\tau}$ of
protons and neutrons, instead of the orbital ones.

The annihilation operator for the $K=1$ scissors mode,  
${\cal O}(\omega_1, K=1) = 
{\cal O}_J(\omega_1, K=1) + {\cal O}_Q(\omega_1, K=1)$, 
is obtained by replacing 
$J_{\tau}^1 \rightarrow - J_{\tau}^{-1}$ and $Q_{\tau}^1 \rightarrow - Q_{\tau}^{-1}$
in (\ref{oj}) and (\ref{oq}). The creation and annihilation operators for the
$K=-1$ modes can then be expressed in terms of the operators for $K=1$ by  
${\cal O}^{\dagger}(\omega_1, K=-1) = 
{\cal O}_J(\omega_1, K=1) - {\cal O}_Q(\omega_1, K=1)$,
and ${\cal O}(\omega_1, K=-1) = 
{\cal O}^{\dagger}_J(\omega_1, K=1) - {\cal O}^{\dagger}_Q(\omega_1, K=1)$.
%

The contribution of the scissors mode to the effective Hamiltonian of the system is
given in analogy to the relation (\ref{hr1}) by
\begin{align}
H_{\rm rot}(\omega_1) 
&= \omega_1 \sum_{K=\pm 1} {\cal O}^{\dagger}(\omega_1, K) {\cal O}(\omega_1, K) 
\label{hrsc} 
\end{align}
Inserting here the forms of the operators, we see that the
effective Hamiltonian has a 
rotational part and a vibrational part, which are given by
\begin{align}
H_{\rm rot}(\omega_1) &=  
\frac{1}{2 I^J} \left\{ \frac{I^J_n}{I^J_p} 
\left[ (J_p^x)^2 + (J_p^y)^2 \right] + \frac{I^J_p}{I^J_n} 
\left[(J_n^x)^2 + (J_n^y)^2 \right] \right.   \no \\ 
&\left. - 2 \left(J_p^x J_n^x + J_p^y J_n^y \right) \right\},
\label{hrot1} \\
H_Q(\omega_1) &= 
- \frac{\chi_{pn} \left(I^{J}\right)^2}{4 \beta_p \beta_n I_p^J I_n^J}
\left( \frac{2 I^J_n}{I^J} \beta_n Q_p^1 - \frac{2 I^J_p}{I^J} \beta_p Q_n^1 \right)^{\dagger}  \no \\
&\times \left( \frac{2 I^J_n}{I^J} \beta_n Q_p^1 - \frac{2 I^J_p}{I^J} \beta_p Q_n^1 \right).
\label{hq}
\end{align}
%

The term (\ref{hq}) represents the restoring force which acts against the proton-neutron 
oscillations.
If one adds the contribution of the
Goldstone mode (\ref{hr2}) to that of the scissors mode (\ref{hrot1}),
one obtains the kinetic part of the 2-rotor model:
\begin{align}
H_{\rm rot} \equiv H_{\rm rot}(\omega_0) + H_{\rm rot}(\omega_1)  \no \\ 
= \frac{\left(J_p^x\right)^2 + \left(J_p^y\right)^2}{2 I^J_p} + 
\frac{\left(J_n^x\right)^2 + \left(J_n^y\right)^2}{2 I^J_n} \,.  
\label{tworot}
\end{align}
This is analogous to the result derived in Ref.\cite{BAERW}, but   
now the 2-rotor Hamiltonian is expressed in terms of
the total proton and neutron angular momentum operators instead of the orbital ones.
%

\section{Formulae needed for the evaluation of correlators in deformed nuclei}
\setcounter{equation}{0}

For the numerical evaluation of the non-interacting correlators, 
one may use the Nilsson model\cite{NI} for single particle states in a deformed potential
with axial symmetry.   
Assigning the ``asymptotic quantum numbers''\cite{BM} $N, n_3, \Lambda, \Omega$ to 
a single particle state $|a \rangle$, 
one can expand it in a spherical
basis with quantum numbers $N, \ell, m, \Omega$ as follows:
\begin{eqnarray}
\lefteqn{|a \rangle \equiv |N n_3 \Lambda \Omega \rangle} \nonumber \\
& & \hspace{-1cm} 
= \sum_{\ell} \left[c^a(\ell, m_-) |N \ell m_- \Omega \rangle +
c^a(\ell, m_+) |N \ell m_+ \Omega \rangle \right]\,,   \nonumber \\ 
\label{expand}
\end{eqnarray}
where ${\displaystyle m_{\mp} = \Omega \mp \frac{1}{2}}$ corresponds to spin up
or down. The coefficients $c^a(\ell, m)$
must be obtained by diagonalizing the mean field Hamiltonian in the spherical basis, and
an example for the results is given in Table 5-2b of Ref.\cite{BM}. 
In the phase convention used in this paper and in Ref.\cite{BM}, the coefficients
$c^a(\ell, m)$ are real and obey the 
orthonormalization relation
\begin{align}
\sum_{\ell} \left( c^a(\ell, m_-) c^b(\ell, m_-) +  
c^a(\ell, m_+) c^b(\ell, m_+) \right) = \delta^{ab} \,.   
\label{on}
\end{align}
The matrix elements of $L^1 = - (L^x + i L_y)/{\sqrt{2}}$ and $S^1 = - (S^x + i S^y)/{\sqrt{2}}$ 
are given in the spherical basis by 
the well known expressions of elementary quantum mechanics, and are also real.
We then obtain the following results:
\begin{eqnarray}
\lefteqn{\langle \alpha|L^1|i \rangle = 
\left(\frac{-1}{\sqrt{2}}\right) \sum_{\ell}} \nonumber \\
& & \hspace{-2cm} \times \left(
\sqrt{\ell(\ell+1) - m_-\,m_+} \, 
c^{\alpha}(\ell, m_+) \, c^{i}(\ell,m_-) \right.   \nonumber \\
& & \hspace{-2cm}  
\left. + \sqrt{\ell(\ell+1) - m_+\,(m_++1)} \, 
c^{\alpha}(\ell, m_++1) \, c^{i}(\ell,m_+)  \right) \,,
\label{l1} 
\end{eqnarray}
\begin{eqnarray}
\langle \alpha|S^1|i \rangle = 
\left(\frac{-1}{\sqrt{2}}\right) \sum_{\ell}  
c^{\alpha}(\ell,m_+) \, c^{i}(\ell,m_+) \,. 
\label{s1}
\end{eqnarray}
In Eqs. (\ref{l1}) and (\ref{s1}), $m_{\pm}$ is defined as ${\displaystyle m_{\pm} = \Omega^i \pm \frac{1}{2}}$, and
we made use of the angular momentum conservation $\Omega^{\alpha} = \Omega^i + 1$.

The above expressions can be used to calculate the $\omega=0$ polarizations
needed for the IEW sum rule, see (\ref{ingmpn}) and (\ref{splitb}). 
The particle states ($\alpha$) and hole states ($i$), which contribute to the correlators, 
depend on the deformation, the number of protons and neutrons, and model assumptions for
the single particle Hamiltonian, like the strength of the $L^2$ term or the spin-orbit
interaction. Therefore the correlators must be calculated in the Nilsson model for each 
nucleus separately.

\section{Sum rules involving spin operators}
\setcounter{equation}{0}

In this Appendix we first briefly explain the derivation of the IEW sum rule
$S_{\rm IEW}$ given in (\ref{scsg}) and (\ref{ss}), before turning to the derivation of 
the EW sum rule $S_{\rm EW}$ given in (\ref{ewfin}).  

\subsection{IEW sum rule (\ref{scsg}) and (\ref{ss})}

We consider the $\omega=0$ limit of the correlator $\Pi^{MM}$ of (\ref{pimm}),
using the form (\ref{simple}). As explained in the main text below Eq.(\ref{iso}),
we can replace the orbital $g$-factors by their isovector parts, 
$g_{\tau} \rightarrow t_{\tau} g_1$. This gives
\begin{eqnarray}
\Pi^{MM}(0) &=& \sum_{\lambda \tau} \left( g_1^2 \, t_{\tau} \Pi^{JJ}_{\tau \lambda}(0) t_{\lambda}
+ g_1 \, t_{\tau} \Pi^{JS}_{\tau \lambda}(0)\,h_{\lambda}  \right. 
\nonumber \\
& &\left. +
g_1 \, h_{\tau} \Pi^{SJ}_{\tau \lambda}(0) + h_{\tau} \Pi^{SS}_{\tau \lambda}(0) h_{\lambda} \right) \,.
\label{ins}
\end{eqnarray}
Using the form (\ref{simple}) and the definitions
$I^J_{\tau} = \pi^{JJ}_{\tau}(0)$, $I^M_{\tau} = \pi^{JS}_{\tau}(0)$,
$I^S_{\tau} = \pi^{SS}_{\tau}(0)$, we easily obtain for the individual terms in (\ref{ins}):
\begin{eqnarray}
\sum_{\tau \lambda} t_{\tau} \Pi^{JJ}_{\tau \lambda}(0) &=& 
\left(I^{J}_p + I^{J}_n \right) - 
\frac{\left(I^{J}_p - I^{J}_n \right)^2}{I^J}
= \frac{4 I^J_p I^J_n}{I^J} \,, \nonumber \\
\label{eins} 
\end{eqnarray}
\begin{align}
&\sum_{\tau \lambda} \left(t_{\tau} \Pi^{JS}_{\tau \lambda}(0) h_{\lambda} +
h_{\tau} \Pi^{SJ}_{\tau \lambda}(0) t_{\lambda} \right)
= 2 \left( h_p {\hat I}^M_p +  h_n {\hat I}^M_n \right)  \nonumber \\
&- \frac{2 \left(I^J_p - I^J_n \right)}{I^J} 
\left( h_p {\hat I}^M_p +  h_n {\hat I}^M_n \right)
= 4 \left(h_p {\hat I}^M_p I^J_n - h_n {\hat I}^M_n I^J_p \right)\,, 
\label{zwei}
\end{align}   
where ${\hat I}^M_{\tau}$ means the real part of $I^M_{\tau}$, and
\begin{eqnarray}
\sum_{\tau \lambda} h_{\tau} \Pi^{SS}_{\tau \lambda}(0) h_{\lambda}
= h_p^2 I^S_p + h_n^2 I^S_n - \frac{|h_p I^M_p + h_n I^M_n|^2}{I^J}\,.
\nonumber \\
\label{drei}
\end{eqnarray} 
Using these forms in (\ref{ins}), we arrive at (\ref{scsg}) and (\ref{ss})
of the main text.

\subsection{EW sum rule (\ref{ewfin})}

We first briefly recapitulate the evaluation of (\ref{ewfree}) and (\ref{ewint}) for the 
case $R'=R=J$. We make use of the identity (\ref{rel}) to express the result in terms
of the $QQ$ bubble graph $\pi^{QQ}_{\tau}(0)$, and the self consistency relation (\ref{gap})
for $\varepsilon_{\tau}=0$ 
to express the final result in terms of the quadrupole field $\langle Q_{\tau}^0 \rangle$:
\begin{eqnarray}
{\lefteqn - \lim_{\omega\rightarrow \infty} \omega^2 \Pi^{JJ}_{\tau \lambda}(\omega)} 
\nonumber \\
& & \hspace{-3cm} = 3 \delta_{\tau \lambda} \beta_{\tau}^2 \pi_{\tau}^{QQ}(0) +
3 \left(\beta_{\tau} \pi_{\tau}^{QQ}(0)\right) \chi_{\tau \lambda} 
\left(\beta_{\lambda} \pi_{\lambda}^{QQ}(0)\right)  
\nonumber  \\
& & \hspace{-3cm}= 3 \delta_{\tau \lambda} \beta_{\tau} \langle Q_{\tau}^0 \rangle  +
3 \langle Q_{\tau}^0 \rangle \chi_{\tau \lambda} \langle Q_{\lambda}^0 \rangle \,. 
\label{jj1}
\end{eqnarray}
Because of the self consistency relation (\ref{gap}) for exact symmetry, it is clear that
this expression vanishes if we sum over $\tau$ or $\lambda$. Therefore we can replace
the orbital $g$-factors in the first term of (\ref{pimm}) by their isovector parts,
i.e., $g_{\tau} \rightarrow t_{\tau} g_1$, and $g_{\lambda} \rightarrow t_{\lambda} g_1$.
The result can again be simplified by using the relation (\ref{gap}), and becomes finally
\begin{align}
- \lim_{\omega\rightarrow \infty} \omega^2 \sum_{{\tau}{\lambda}} \left(g_{\tau} 
\Pi^{JJ}_{\tau \lambda}(\omega) g_{\lambda}\right) =  
-12 g_1^2 \langle Q_p \rangle \chi_{pn} \langle Q_{n}^0 \rangle \,. 
\no \\ 
\label{jj2}
\end{align}  
Next we will show that the second and third terms in the correlator (\ref{pimm}) vanish
in the limit which is needed for the EW sum rule of (\ref{ew}). For this purpose, we need
the counterparts of some of the identities in the main text for the spin operator.
The identities analogous to (\ref{comm0}) and (\ref{rel}) are
\begin{eqnarray}
\left[H_{0 \tau}, S_{\tau}^{\pm 1} \right] &=& \sqrt{2} \gamma_{\tau} V_{\tau}^{\pm 1}  
\label{spin1} \\
\langle \alpha| S_{\tau}^{\pm 1} |i \rangle &=& \frac{\sqrt{2}\gamma_{\tau}}{\omega_{\alpha i}}
\langle \alpha |V_{\tau}^{\pm 1} |i \rangle \,.  
\label{spin2}
\end{eqnarray}
Here $\gamma_{\tau}$ is the strength parameter of the spin-orbit interaction (see Eq.(\ref{h0})), 
and we defined the 
operator $V^K$ as the tensor product of order $1$ of the orbital and spin angular momentum operators. 
In the notation of first quantization,
\begin{align}
V^K = \left[ L \times S \right]_{(1)}^K = \frac{i}{\sqrt{2}} \left( \vec{L} \times \vec{S} \right)^K\,,
\label{v}
\end{align}
where the product in the last expression denotes the usual vector product.
Then, without making use of any symmetry constraints, the following identity for
the correlator $\Pi_{\tau \lambda}^{WS}$ with one arbitrary operator ($K=1$ component $W^1$) and the
spin operator $S^1$ can be derived (compare to the relation (\ref{wt1}) in the main text):
\begin{align}
\omega \, \Pi_{\tau \lambda}^{WS}(\omega) =  \sqrt{2}\, 
\Pi_{\tau \lambda}^{WV} (\omega) \gamma_{\lambda} 
+ \delta_{\tau \lambda} \langle \left[ S_{\tau}^1, W_{\tau}^{1 \dagger} \right]  \rangle\, . 
\label{wts}
\end{align}
By setting $\omega=0$ in this relation, we obtain the identity (compare to (\ref{w1}))
\begin{align}
\Pi_{\tau \lambda}^{WV}(0) \gamma_{\lambda} = - \frac{1}{\sqrt{2}} \delta_{\tau \lambda}
\langle \left[ S_{\tau}^1, W_{\tau}^{1 \dagger} \right]  \rangle\, .
\label{gs}
\end{align}
In particular, for the case $W=Q$ this identity implies that
$\Pi_{\tau \lambda}^{QV}(0)  = 0$, because the quadrupole operator obviously commutes with the
spin operator.
By using the RPA equation (\ref{rpag1}) for $W' = Q$ and $W = V$, we see that also the non-interacting
correlator vanishes for $\omega=0$, i.e.,
\begin{align}
\Pi_{\tau \lambda}^{QV}(0) = \pi_{\tau}^{QV}(0) = 0 \,.
\label{vanish}
\end{align}
Returning now to the evaluation of (\ref{ewfree}) and (\ref{ewint}) of the main text for the case 
$R'=J$ and $R=S$,
we obtain from the identities (\ref{rel}) and (\ref{spin2})
\begin{eqnarray}
\lefteqn{ - \lim_{\omega \rightarrow \infty} \omega^2 \Pi_{\tau \lambda}^{JS}(\omega)
 = - \sqrt{6} \delta_{\tau \lambda} \beta_{\tau} \gamma_{\tau} \pi_{\tau}^{QV}(0)}
\nonumber \\
& & \vspace{-3cm} 
- \sqrt{6} \left(\beta_{\tau} \pi_{\tau}^{QQ}(0) \right) \chi_{\tau \lambda} 
\left(\gamma_{\lambda} \pi_{\lambda}^{QV}(0) \right) \nonumber \\
& & \vspace{-3cm} = - \sqrt{6} \delta_{\tau \lambda} \beta_{\tau} \gamma_{\tau} \pi_{\tau}^{QV}(0) -
\sqrt{6} \left( Q_{\tau}^0 \right) \chi_{\tau \lambda} 
\left(\gamma_{\lambda} \pi_{\lambda}^{QV}(0) \right) 
\nonumber \\
& & \vspace{-3cm} = 0 \,, 
\label{jsew} 
\end{eqnarray}
where we used (\ref{vanish}) in the last step. We therefore conclude that there are no
$J-S$ crossing contributions, corresponding to the second and third terms in (\ref{pimm}), 
to the EW sum rule (\ref{ew}).

Turning finally to the $S-S$ contribution to the EW sum rule, we obtain for the case
$R'=R=S$ in (\ref{ewfree}) and (\ref{ewint})
\begin{eqnarray}
\lefteqn{
- \lim_{\omega \rightarrow \infty} \omega^2 \Pi_{\tau \lambda}^{SS}(\omega) 
= 2 \delta_{\tau \lambda} \gamma_{\tau}^2 \pi_{\tau}^{VV}(0)}  
\nonumber \\
& & \hspace{-1cm} 
+ 2 \left( \gamma_{\tau} \pi_{\tau}^{VQ}(0) \right) \chi_{\tau \lambda}
\left( \gamma_{\lambda} \pi_{\lambda}^{QV}(0) \right)  
= 2 \delta_{\tau \lambda} \gamma_{\tau}^2 \pi_{\tau}^{VV}(0) \,,   \nonumber \\
\label{ssew}
\end{eqnarray}
where we used the result (\ref{vanish}) in the last step. We therefore see that in our
present schematic model, which does not include spin-spin interactions, the $S-S$
correlator in the limit $\omega \rightarrow \infty$ is simply the non-interacting one. 
In order to express $\pi_{\tau}^{VV}(0)$
in terms of a ground state expectation value, we use the identity (\ref{gs}) for the
case $W=V$. The commutator on the r.h.s. can be decomposed into tensors of rank $0$,
$1$ and $2$. Because only the scalar term can have a ground state expectation value,
the identity (\ref{gs}) gives
\begin{align}
\Pi_{\tau \lambda}^{VV}(0) \gamma_{\lambda} = \frac{1}{3} \, \delta_{\tau \lambda} 
\langle {\bold L}_{\tau} \cdot {\bold S}_{\tau} \,. \rangle
\label{pivv}
\end{align}
From the RPA equation (\ref{rpag1}) for $W'=W=V$ and the result (\ref{vanish}), we see that
$\Pi_{\tau \lambda}^{VV}(0) = \delta_{\tau \lambda} \pi^{VV}(0)$, and we obtain from (\ref{ssew}) 
and (\ref{pivv})
\begin{align}
- \lim_{\omega \rightarrow \infty} \omega^2 \Pi_{\tau \lambda}^{SS}(\omega) 
= \frac{2}{3} \delta_{\tau \lambda} \gamma_{\tau} 
\langle {\bold L}_{\tau} \cdot {\bold S}_{\tau} \rangle \,.
\label{fins}
\end{align}
Using this result in the $S-S$ term of the EW sum rule, which arises from the last term in
(\ref{pimm}), we obtain the result given in Eq.(\ref{ewfin}) of the main text.

It is clear from the identity (\ref{pivv}) that 
$\langle {\bold L}_{\tau} \cdot {\bold S}_{\tau} \rangle > 0$, because the  
correlator $\Pi_{\tau \lambda}^{VV}(0) = \delta_{\tau \lambda} \pi_{\tau}^{VV}(0)$ is positive due to
the spectral representation (\ref{pi}). 
It is also easy to see that this term
is nothing but the contribution from the mean field Hamiltonian ($H_0$) to
the familiar double commutator, i.e.
\begin{align}
\langle 0| \left[ \left[ H_0, M^1 \right], M^{-1} \right] | 0 \rangle
=  \frac{2}{3} \sum_{\tau} h_{\tau}^2 \gamma_{\tau} \langle {\bold L}_{\tau} \cdot
{\bold S}_{\tau} \rangle \,. 
\label{ewh0}
\end{align}
In the diagrammatic approach used in the main text of this paper,
this is reflected by the fact that 
the second term in (\ref{ewfin}) comes from the non-interacting
correlator $\pi^{SS}$, see in particular Eq.(\ref{ssew}).  


We finally add a remark on Eq.(\ref{vanish}): Because the operators $Q$ and $V$ have the same 
$T$-symmetry (both $T$-even),
the identity $\Pi_{\tau \lambda}^{QV}(0)  = 0$ does not follow from the spectral representation
(\ref{spec1}). Also, this identity does not rely on rotational symmetry, i.e., it holds
also for finite symmetry breaking parameters $\varepsilon_{\lambda}$. Therefore, if 
we use $\Pi_{\tau \lambda}^{VQ}(0)  = 0$ on the l.h.s. of the
identity (\ref{w1}) of the main text for finite $\varepsilon_{\lambda}$, we can confirm that
$\langle V_{\tau}^0 \rangle =0$.


\begin{thebibliography}{99}

\bibitem{SC} D. Bohle, A. Richter, W. Steffen, A.E.L. Dieperink,
N. Lo Iudice, F. Palumbo, and O. Scholten, Phys. Lett. {\bf B 137}
(1984) 27.
%
\bibitem{NCIM} N. Lo Iudice, Rivista Nuovo Cimento {\bf 9} (2000) 1. 
%
\bibitem{TWOROT} N. Lo Iudice, and F. Palumbo, Phys. Rev. Lett. {\bf 41}
(1978) 1532.
%
\bibitem{FI} F. Iachello, Nucl. Phys. {\bf A 358} (1981) 89c.
%
\bibitem{SCREV} K. Heyde, P. von Neumann-Cosel, and A. Richter,
Rev. Mod. Phys. {\bf 82} (2010) 2365.
%
\bibitem{IU1} N. Lo Iudice, Nucl. Phys. {\bf A 605} (1996) 61.
%
\bibitem{ZA} D. Zawischa, J. Phys. G {\bf 24} (1998)
683.
%
\bibitem{BM} A. Bohr, and B.R. Mottelson, {\em Nuclear Structure}, Vol. II 
(World Scientific, 1998).
%
\bibitem{BAERW} W. Bentz, A. Arima, J. Enders, A. Richter, J. Wambach,
Phys. Rev. {\bf 84} (2011) 014327.
%
\bibitem{UTFU} H. Ui and G. Takeda, Prog. Theor. Phys. {\bf 70} (1983) 176; \\
K. Fujikawa and H. Ui, Prog. Theor. Phys. {\bf 75} (1986) 997.
%
\bibitem{PW} T. Papenbrock, Nucl. Phys. {\bf A 852} (2011) 36; \\
T. Papenbrock and H. A. Weidenm\"uller, arXiv:1307.1181 [nucl-th].
%
\bibitem{SRM} E. Lipparini, and S. Stringari, Phys. Lett. {\bf B 130} (1983) 139.
%
\bibitem{LS} E. Lipparini, and S. Stringari, Phys. Rept. {\bf 175} (1989) 103.
%
\bibitem{PRC57} N. Lo Iudice, Phys. Rev. {\bf C 57} (1998) 1246. 
%
\bibitem{PRC71} J. Enders, P. von Neumann-Cosel, C. Rangacharyulu,
and A. Richter, Phys. Rev. {\bf C 71} (2005) 014306.
%
\bibitem{ZMS} D. Zawischa, M. Macfarlane, and J. Speth, Phys. Rev. {\bf C 42} 
(1990) 1461.
%
\bibitem{NEW} W. Bentz, A. Arima, A. Richter, and J. Wambach,
{\em Analytic approach to nuclear rotational states: The role of spin
- A model including pairing and spin dependent interactions-} (in preparation).
%
\bibitem{RO} D.J. Rowe, Nuclear Collective Motion,
     Methuen and Co., 1970.
%
\bibitem{MES} A. Messiah, Quantum Mechanics (Dover Publications, 1999),
Appendix C. 
%
\bibitem{ING} D.R. Inglis, Phys. Rev. {\bf 96} (1954) 1059.
%
\bibitem{YT} Y. Takahashi, Nuovo Cim. {\bf 6} (1957) 371.
%
\bibitem{MIG} A.B. Migdal, {\em Theory of finite Fermi systems and
applications to atomic nuclei}, Wiley, New York, 1967.
%
\bibitem{NI} S.G. Nilsson, Mat. Fys. Medd. Dan. Vid. Selsk. {\bf 29}, No.16
(1955) 1.
%
\bibitem{NA} Y. Nambu, Physica {\bf 15 D} (1985) 147.
%
\bibitem{MROT} S. Frauendorf, Rev. Mod. Phys. {\bf 82} (2010) 2365; \\ 
L.F. Yu, P.W. Zhao, S.Q. Zhang, P. Ring, J. Meng, Phys. Rev. {\bf C 85} (2012) 024318.
%
\bibitem{TWO}  W. Bentz,
{\em Analytic approach to nuclear rotational states: The role of spin - A model
with dynamical breaking of spin symmetry - } (in preparation).
%
\bibitem{RS} P. Ring, and P. Schuck, The Nuclear Many-Body Problem,
Springer, 1980.
%
\bibitem{CL}   See for example: K. Shimizu, Z. Physik {\bf 278} (1976) 201.

\end{thebibliography}
\end{document}